\documentstyle[aps,prd]{revtex}

\input epsf

\begin{document}

\preprint{DUKE-TH-96-99}

\title{Classical Fields Near Thermal Equilibrium}

\author{Carsten Greiner$^{1}$ and Berndt M\"uller$^{2}$}

\address{$^{1}$Institut f\"ur Theoretische Physik, Justus-Liebig-Universit\"at
Giessen,\\
D-35392 Giessen, Germany,
\\[2mm]
$^{2}$Department of Physics, Duke University,\\
Durham, North Carolina, 27708-0305, USA.}

\maketitle

\date{DRAFT: \today}

\begin{abstract}
We discuss the classical limit for the long-distance (``soft'') modes
of a quantum field when the hard modes of the field are in thermal
equilibrium.  We address the question of the correct semiclassical
dynamics when a momentum cut-off $\vert{\bf p}\vert \le k_c \ll T$ is
introduced.
Higher order contributions leads to a stochastic interpretation
for the effective action in analogy to Quantum Brownian Motion,
resulting in dissipation and decoherence for the evolution of the soft modes.
Particular emphasis is put on the understanding of dissipation.
Our discussion focuses mostly on scalar fields, but we
make some remarks on the extension to gauge theories.
\end{abstract}


\section{Introduction}

Solutions of the classical field equations in Minkowski space have
been widely used in recent years to describe long-distance properties
of quantum fields that require a nonperturbative analysis.  These
applications include: the diffusion rate of the topological charge in
the electroweak gauge theory \cite{1,2,3,4,5}, the thermalization rate
of non-Abelian gauge fields \cite{6,7,8}, as well as a wide range of
cosmological problems, such as thermalization, decoherence, and
structure formation \cite{9,10,11,12,13,14,15,16}.  Many of these
studies are concerned with dynamical properties of quantum fields near
thermal equilibrium.

Classical treatments of the long-distance dynamics of bosonic quantum
fields at high temperature are based on the observation that the average 
thermal amplitude of low-momentum modes is large.  For a weakly coupled 
quantum field the occupation number of a mode with wave vector {\bf p}
and frequency $\omega_{\bf p}$ is given by the Bose distribution
\begin{equation}
n(\omega_{\bf p}) = \left(e^{\hbar\omega_{\bf p}/T} -1 \right)^{-1}.
\label{Bose}
\end{equation}
For temperatures $T$ much higher than the (dynamical) mass scale $m^*$
of the quantum field, the occupation number becomes large and
approaches the classical equipartition limit
\begin{equation}
n(\omega_{\bf p}) \buildrel \vert {\bf p}\vert \to 0 \over \longrightarrow
{T\over m^*} \gg 1. \label{climit}
\end{equation}
The classical field equations should provide a good approximation for
the dynamics of such highly occupied modes.\footnote{Note that for
interacting quantum fields, generally, the dynamical mass gap at
finite temperature is of order $gT$, where $g$ is a dimensionless
coupling constant.  Hence, the validity of (\ref{climit}) requires
weak coupling. On the other hand, rapid thermalization and decoherence
of the soft modes requires that the coupling to the heat bath is
sufficiently strong \cite{S82}. It needs to be established whether
both conditions can be satisfied simultaneously.}

At a closer look, however, the cogency of this heuristic argument
suffers considerably.  The thermodynamics of a classical 
field\footnote{Indeed, the ultraviolet divergence of thermodynamic 
quantities was one of the motivations for Planck's conjecture in 1900 
that the excitations of field modes are quantized.} is only defined if 
an ultraviolet cut-off $k_c$ is imposed on the momentum {\bf p} such as a
finite lattice spacing $a$.  Many, if not most, thermodynamical 
properties of the classical field depend strongly on the value of the 
cut-off parameter $k_c$ and diverge in the continuum limit $(k_c \to\infty)$.
Examples are the energy density
\begin{equation}
\epsilon(T) \sim \int \frac{d^3p}{(2\pi \hbar)^3}\;
\hbar\omega_{\bf p} n(\omega_{\bf p})
\sim Tk_c^3
\label{edensity}
\end{equation}
and the dynamical mass gap
\begin{equation}
m^{*2}\sim \frac{g^2}{2} \int \frac{d^3p}{(2\pi )^3}\;
{1\over\omega_{\bf p}} n(\omega_{\bf p}) \sim \frac{g^2}{2}T \hbar k_c
\, \, \, ,
\label{mgap}
\end{equation}
in the high-temperature limit
(and for $k_c \gg m^*$), which contrasts with the results
$\epsilon(T)\sim T^4$ and $m^{*2} \sim g^2T^2$ in the full quantum
theory.  Remarkably, some thermodynamic quantities, such as the
diffusion rate of the topological charge \cite{2,5} in non-Abelian gauge
theories or the maximal Lyapunov exponent \cite{6}, which is equivalent 
to the damping rate of soft thermal excitations \cite{8}, are found
to be insensitive to the cut-off, attaining a finite value in the
continuum limit.

The question arises whether the value reached in the continuum limit 
agrees with the one obtained in the full thermal quantum field theory, 
if a complete nonperturbative calculation were feasible.  Naively, one 
expects that any quantity that is independent of $\hbar$ in the lowest 
order of its coupling constant expansion---such as quantities that 
contain only the combination $g^2T$ in thermal non-Abelian gauge 
theories---should be reliably calculable in the classical limit.  Of 
course, there will exist quantum corrections which are suppressed by 
powers of the coupling constant $\alpha=g^2\hbar$, i.e. such a quantity 
will have a perturbative expansion of the form
\begin{equation}
(g^2T)^n (c_0+c_1\alpha+c_2\alpha^2+ \ldots) \label{form}
\end{equation}
where only the coefficient $c_0$ is calculable classically, but the
coefficients $c_i\; (i\ge 1)$ can be obtained using perturbative
techniques.

Other questions also arise:  What coupling constant $g$ should be used
in the classical calculation?  How is it renormalized?  How does mass
renormalization enter into the equation for the low-momentum modes
that are described classically?  How does the classical equation
``know'' that it derives its justification from the classical limit of
a thermal quantum field?  How does the temperature enter into the
classical equation?  Which mechanism ensures that the attained temperature 
equals that of the underlying thermal field theory, if the 
classically described modes thermalize due to their interaction?

Some of these questions were addressed in recent articles
\cite{12,17,Mo90,18}; others remain unanswered.  Here we make an attempt to
address the complex of questions in its entirety, applying the method
of effective field theories and the real-time formulation of finite
temperature quantum field theory.
Our approach leads us to the formulation of an effective dynamics
for the soft modes of the quantum field in terms of a stochastic,
dissipative action. The stochastic forces describe the exchange of
energy and other quantum numbers with the perturbatively treated
hard modes, and the dissipative terms describe the eventual approach
to thermal equilibrium of the soft modes. A fluctuation-dissipation
theorem ensures that the soft modes reach the same temperature as
the hard modes with which they interact.
We perform our analysis for the simplest nontrivial example:  a massive
scalar field with quartic self-interaction in ($3+1$) dimensions.
This allows us to evade the technical difficulties associated with
gauge invariance and to make contact with previous work
\cite{9,12,17,Mo90,18}.

It is worthwhile mentioning that a similar issue arises in the Euclidean 
formulation of thermal non-Abelian gauge theories, because thermodynamic 
quantities, such as the pressure or screening masses,
receive contributions at a scale $(g^2T)^n$, which are essentially 
classical and usually cannot be calculated by means of perturbative 
methods.  Recently, Braaten and Nieto \cite{B94,BN95} have shown how 
low-energy effective actions in Euclidean space can be utilized to 
systematically resolve this problem.  The analogous problem in Minkowski 
space is more complicated, since dimensional reduction \cite{AP81}
does not occur in this case because at finite temperature there is no
gap in the excitation spectrum.  In this case the effective long-distance 
field theory becomes essentially classical, because quantal loop
corrections are suppressed by powers of $\hbar k_c/T\ll 1$ where $k_c$
is the ultraviolet cut-off of the effective theory.  However, the
dynamics retains its full $(3+1)$ dimensionality.

Our manuscript is organized as follows: In the next Section we
first review the concepts of effective actions and influence functionals.
We then derive the influence functional for a self-coupled scalar
quantum field and obtain the classical, stochastic equations of motion
for the soft field modes. In Section III we show in detail how dissipation 
emerges from the effective equation of motion and how the proper
Markovian limit of the corresponding nonlocal terms can be obtained.  
Section IV briefly considers the generalization of our results to gauge 
theories and presents a discussion of some open issues.

\section{Real-time Effective Action for Classical Soft Modes}

\subsection{Preliminaries}

Our goal is the derivation of an effective action for the classical
dynamics in Minkowski space of the low-momentum modes of a scalar quantum 
field $\Phi$ near thermal equilibrium.  Following B\"odeker, McLerran, 
and Smilga \cite{17} and Lombardo and Mazzitelli \cite{18}, we divide the 
Fourier components of the field $\Phi$ into low-momentum modes with $\vert
{\bf p}\vert \le k_c$ (henceforth called ``soft'' modes) and
high-momentum modes with $\vert {\bf p}\vert >k_c$ (called ``hard''
modes from now on) by defining\footnote{In the notation of ref. \cite{17}:
$\phi=\Phi_{\rm L},\; \varphi=\Phi_{\rm S}$; in that of ref. \cite{18}:
$\phi=\Phi_<, \; \varphi=\Phi_>$.}
\begin{equation}
\Phi({\bf p},t) = \Phi({\bf p},t)\theta (k_c-\vert{\bf p}\vert) +
\Phi({\bf p},t)\theta(\vert{\bf p}\vert-k_c) \equiv \phi({\bf p},t) +
\varphi({\bf p},t). \label{hmode}
\end{equation}
It is clear from the introduction that the cutoff scale $k_c$ should be
much smaller than the temperature T, i.e. $k_c \ll T$. Still $k_c$ may be
chosen comparable with the dynamical mass scale $m^* \sim gT$ for the 
quasi-classical regime.  We assume that all hard modes $\varphi$ are, and 
remain, thermally occupied.\footnote{This is, potentially, a weak point 
because the thermalization rate of hard modes is usually not significantly 
larger than the one of soft modes. By using a general form of the density
matrix for the hard modes, the assumption of their thermalization could, 
in principle, be relaxed.}  We do not make this assumption for the soft 
modes $\phi$, but we assume that their occupation number is sufficiently 
large to warrant a classical treatment.

We note here that this procedure is not the same as obtaining an
``effective action'' (in the common usage of the expression) for the
expectation values of the quantum field in the presence of classical
sources.  The latter entails calculating the generating functional
\begin{equation}
Z(J) = \int {\cal D}\Phi\; e^{i(S[\Phi]+J\phi)} \equiv e^{iW[J]}
\label{gfunct}
\end{equation}
allowing for sources $J$ of the soft modes only, and then taking the
Legendre transform
\begin{equation}
\Gamma (\bar\phi) = W[J] -J {\delta W\over \delta J} 
\label{Ltransf}
\end{equation}
with $\bar\phi  = \langle \phi \rangle _J = \delta W/\delta J$.
The effective action $\Gamma
(\bar\phi)$ contains all quantal and thermal corrections to 
the field expectation values.  

Our definition of an ``effective action'' for the soft modes 
is\footnote{Below we will generalize the effective action to the situation
appropriate for real-time evolution, which requires a doubling of the
degrees of freedom:
$S_{\rm eff}^{k_c}[\phi] \to S_{\rm IF}^{k_c}[\phi,\phi']$.}
\begin{equation}
e^{iS_{\rm eff}^{k_c}[\phi]} = \int {\cal D}\varphi\; e^{iS[\Phi]} \equiv
\int {\cal D}\varphi\; e^{iS[\phi,\varphi]}, \label{effective}
\end{equation}
which excludes all quantal and thermal corrections from the soft modes
themselves.  The two definitions only agree in the limit $k_c\to 0$:
\begin{equation}
S_{\rm eff}^{k_c} [\phi] \buildrel k_c\to 0 \over \longrightarrow
\Gamma [\langle \phi \rangle ].  \label{dagree}
\end{equation}

Which approach is preferable depends on the infrared behavior of the
quantum field under consideration.  When it is well behaved, so that
perturbation theory---possibly after resummation of a certain class of
diagrams---provides an adequate description of the low-momentum sector
of the field, the standard effective action $\Gamma(\phi)$ is the best
choice.  $\Gamma(\phi)$ already accounts for all thermal and quantal
corrections and determines, in addition to the evolution of the field
expectation value, all $n$-point functions of the field.  On the other
hand, if the low-momentum field modes are governed by truly
nonperturbative physics, such as in the case of non-Abelian gauge
theories at high temperature, we have to rely on a numerical treatment
of the long-distance dynamics on the basis of $S_{\rm eff}^{k_c}[\phi]$.
The major advantage then is that all the short-distance dynamics that
can be treated perturbatively is already contained in the effective
action $S_{\rm eff}^{k_c}$,
thus the numerical evaluation can concentrate on the
nonperturbative sector alone where quantal corrections
among the soft modes are of minor
importance while classical (thermal) effects dominate.

The derivation of an effective equation of motion for the soft modes
by elimination of the fast modes corresponds to a coarse graining of
the quantum field.  In the spirit of the general formalism of quantum
statistics, the assumption of thermal equilibrium for the eliminated
modes can be interpreted as one of minimal information about the
microscopic quantum state of the hard modes.  The general theory of
coarse graining \cite{19} then permits several statements about the
nature of the effective dynamics of the soft modes:

\begin{enumerate}
\item The mass parameters and the coupling constants are renormalized, 
and new effective interactions, which are in general nonlocal, will appear.

\item The dynamics is dissipative and contains noise terms 
satisfying a fluctuation-dissipation theorem. The semiclassical real-time 
evolution is inherently stochastic, reflecting the coupling of the
soft modes to the heat reservoir of hard thermal modes.

\item The dissipative terms are nonlocal in time, giving rise to memory 
kernels stretching over the history of the evolution.  This occurs because 
the heat bath needs time to respond.  A change of the system (here the soft 
modes) influences the heat bath which in turn responds causally back onto 
the system.  
\end{enumerate}

The most convenient techniques for the formulation of the problem and
the derivation of the effective action are the closed-time-path (CTP)
method of Schwinger and Keldysh \cite{20} and the influence
functional (IF) method of Feynman and Vernon \cite{21}.  As discussed,
e.g., in ref. \cite{11,Su88} these two approaches yield identical classical
equations for the expectation values of the coarse-grained field components
$\phi$.
In appendix A we summarize the basic idea of the influence action and its
properties for a quantum mechanical system {\sl X} coupled to a heat bath
{\sl Q}.
The influence action $S_{\rm IF}[\phi,\phi';t_f]$ is defined
as (see appendixes A and B)
\begin{equation}
e^{iS_{\rm IF}[\phi,\phi';t_f]} = \int d\varphi_f d\varphi_i
d\varphi'_i \int_{\varphi_i}^{\varphi_f} {\cal D}\varphi
\int_{\varphi'_i}^{\varphi_f} {\cal D}\varphi' \;
e^{i(S[\varphi]+S_{\rm int}[\phi,\varphi]-S[\varphi']-S_{\rm int}
[\phi',\varphi'])} \rho_{\rm h}(\varphi_i,\varphi'_i;t_i)
\label{IF}
\end{equation}
where $\rho_h(\varphi_i,\varphi'_i)$ is the density matrix of initial
configurations of the hard field modes and 
\begin{equation}
S[\Phi] = S[\phi]+S[\varphi]+S_{\rm int}[\phi,\varphi] \label{action}
\end{equation}
is the action for the scalar field.
The influence action vanishes for 
$\phi=\phi'$; its variation with respect to $\phi_{\Delta}=(\phi-\phi')$ 
yields the semiclassical
corrections to the equation of motion for the soft field modes
$\phi$ due to their interaction with the hard modes $\varphi$.

The generating functional in the closed-time-path method is defined by
introducing sources for the soft field modes only:
\begin{equation}
e^{iW_{\rm CTP}[J,J']} = \int d\Phi_f d\Phi_i d\Phi'_i
\int_{\Phi_i}^{\Phi_f} {\cal D}\Phi \int_{\Phi'_i}^{\Phi_f} {\cal
D}\Phi'\; e^{i(S[\Phi]-S[\Phi'] + J\phi-J'\phi')} 
\rho (\Phi_i,\Phi'_i;t_i),
\label{sfmodes}
\end{equation}
where here the limit $t_i\to -\infty$, $t_f \to +\infty$ is understood.
Assuming that the initial density matrix of the full field factorizes
into one for the hard and one for the soft modes,
\begin{equation}
\rho(\Phi_i,\Phi'_i;t_i) = \rho_{\rm h}(\varphi_i,\varphi'_i;t_i)
\otimes \rho_{\rm s}(\phi_i,\phi'_i;t_i),
\label{factor}
\end{equation}
the generating functional can be expressed in terms of the influence
functional as
\begin{equation}e^{iW_{\rm CTP}[J,J']} = \int d\phi_f d\phi_i d\phi'_i
\int_{\phi_i}^{\phi_f} {\cal D}\phi \int_{\phi'_i}^{\phi_f} {\cal
D}\phi'\; e^{i(S[\phi]-S[\phi']+J\phi-J'\phi'+S_{\rm
IF}[\phi,\phi';\infty])} \rho_{\rm s}(\phi_i,\phi'_i;t_i).
\label{gfunctional}
\end{equation}
If the classical approximation can be made here, i.e. if loops
involving soft momenta can be neglected and the density matrix
$\rho_{\rm s}(\phi_i,\phi'_i;t_i)$ can be assumed as diagonal, 
the expectation values of the soft field modes are described by 
the effective closed-time path action
\begin{equation}
\Gamma_{CTP}[\phi,\phi'] \approx S[\phi] - S[\phi'] + S_{\rm IF}
[\phi,\phi';\infty]
\label{ctaction}
\end{equation}
where $\phi=\partial W_{\rm CTP}/\partial J$ and $\phi'=\partial
W_{\rm CTP}/\partial J'$.

The main step to be taken in the derivation of effective classical
equations of motion for the soft field modes are therefore: (1)
calculating the influence action due to the hard modes, and (2)
showing that loop contributions involving soft modes can be neglected.
It is our objective in the next section to derive the influence action
for the long-wavelength modes and then to discuss in the following
sections the meaning of the additional terms contributing to the 
quasi-classical equation of motion for the soft modes in more detail.

\subsection{Influence Action}

In most cases of interest the influence action must be obtained from a 
suitably truncated perturbation expansion.  This raises the question how 
far one has to go in the expansion to obtain all important physical 
contributions.  Here we will concentrate on the case of a self-interacting 
scalar field with action
\begin{equation}S[\Phi] = \int_{t_i}^{t} \, d^4x
\left[ \textstyle{{1\over 2}} \partial_{\mu}\Phi
\partial^{\mu}\Phi - \textstyle{{1\over 2}}m^2\Phi^2 -
\textstyle{{1\over 4!}} g^2\Phi^4 \right],
\label{expansion}
\end{equation}
where $t>t_i$ describes the time of observation. Details of the derivation of
$S_{\rm IF}[\phi,\phi']$ are given in appendix B. It is our purpose here
to outline the strategy and the results obtained.
As will become clear, it turns out that one must
include all diagrams up to order $g^4$ in
the perturbative expansion, even if $g^2\ll 1$.  The interaction that
generates $S_{\rm IF}[\phi,\phi']$ is
\begin{equation}
S_{\rm int}[\phi,\varphi] = - \, \int_{t_i}^{t} \, d^4x
\left[ \textstyle{{1\over 6}} g^2 \left( \phi^3\varphi
+ \textstyle{{3\over 2}}\phi^2\varphi^2 + \phi\varphi^3 \right)
+ \textstyle{{1\over 4!}} g^2\varphi^4 \right].
\label{intera}
\end{equation}

All non-vanishing diagrams contributing to $S_{\rm IF}$ up to second order in
$S_{\rm int}$ are shown in Figure 1. The diagrams (a)--(d) are the
familiar one-particle irreducible diagrams appearing in the expansion
of the effective action $\Gamma (\bar\phi)$.  The fifth diagram (e)
vanishes in the exact limit $k_c\to 0$.  It contributes to the 
coarse-grained effective action whenever $k_c\not= 0$.
Other diagrams not listed vanish due to the constraint of
(3-dimensional) momentum conservation.
The inclusion of
the diagrams of order $g^4$ is crucial, because this is the lowest
order at which the real-time effective action develops an imaginary
part which gives rise to stochastic terms
as well as the corresponding real part will
give rise to dissipative terms in the equations of motion for the soft modes.
We emphasize that this property is not connected
with the number of loops in the diagram:  the two-loop diagram (c) as
well as the one-loop diagram (d) and the tree diagram (e) contribute
to noise and dissipation.  

We define the real-time propagators for the hard modes as
\begin{eqnarray}
G_{++}^{k_c} (x-y) &= &i\langle {\rm T} \left(
\varphi(x)\varphi(y)\right)\rangle
\equiv G_F^{k_c}(x-y) \nonumber \\
G_{+-}^{k_c} (x-y) &= &i\langle\varphi(y)\varphi'(x)\rangle
\equiv G_<^{k_c}(x-y) \nonumber \\
G_{-+}^{k_c} (x-y) &= &i\langle \varphi'(x) \varphi (y)\rangle
\equiv G_>^{k_c}(x-y) \nonumber \\
G_{--}^{k_c} (x-y) &= &i\langle \bar {\rm T} \left( \varphi'(x) \varphi'(y)
\right) \rangle , \label{diagrams}
\end{eqnarray}
where the superscript ``$k_c$'' indicates that these propagators are
defined in the restricted space of momenta $\vert{\bf p}\vert >k_c$.
As discussed in the introduction, we assume that the hard modes are
thermally populated, with a simple dispersion relation
\begin{equation}
\omega_{\bf p} = \sqrt{{\bf p}^2+m^2} \, \, \, .  \label{disrel}
\end{equation}
(At high temperature where the bare mass $m$ can be neglected,
i.e. $m \ll gT$, choosing $k_c$ on the order of the dynamical 
mass scale $gT$ or even smaller, one can modify this dispersion by taking
into account, to lowest order, the dynamically generated mass
(of order $\sim gT$) for the hard modes themselves.)

The complex influence action $S_{\rm IF}[\phi,\phi']$ can be expanded 
in powers of the fields $\phi,\phi'$. Up to order $g^4$, this expansion
has the following structure (see appendix B):
\begin{eqnarray}
{\rm Re}\;S_{\rm IF} &= &-\int_{t_i}^{t} d^4xd^4y \sum_{N=1}^3 
{1\over (2N)!}
\; R_N^-(x)\; \theta (x_0-y_0) {\rm Re} \left[ \Gamma_{2N}^{k_c}(x-y)\right]\;
R_N^+(y) \nonumber \\
{\rm Im}\; S_{\rm IF} &= &-\int_{t_i}^{t} d^4xd^4y \sum_{N=1}^3 
{1\over (2N)!}
\; R_N^-(x)\; {\rm Im}\left[ \Gamma_{2N}^{k_c}(x-y)\right]\;
R_N^-(y),
\label{expanded}
\end{eqnarray}
where $\Gamma_{N}^{k_c}$ is the amputated $2N$-point vertex function,
restricted to contributions solely from the hard modes with $\vert{\bf
p}\vert>k_c$.  Here we have used the abbreviations
\begin{equation}
R_N^{\pm}(x) = \phi(x)^N \pm \phi'(x)^N. \label{abbreviations}
\end{equation}
Up to order $g^4$, the vertex functions $\Gamma_2^{k_c},\;\Gamma_4^{k_c},\;
\Gamma_6^{k_c}$ are given by the diagrams in Figure 1.  All higher vertex
functions vanish.  $\Gamma_2^{k_c}$ is the one-particle
irreducible self energy of the soft field modes as generated by the
hard modes, $\Gamma_4^{k_c}$ is the one-loop correction to the bare
four-particle vertex, and $\Gamma_6^{k_c}$ is a one-particle reducible
six-particle vertex which vanishes in the limit $k_c \to 0$.  Detailed
expressions for the vertex functions are given in
Appendix B.

\subsection{Interpretation of the imaginary part of the Influence Action}

Before one can derive classical equations of motion for the soft field
modes, one has to deal with the imaginary part of the influence
action.  The general idea of its physical significance
is outlined in appendix A.
It has to be interpreted as a stochastic ``external'' force
driving the classical degrees of freedom, i.e.~the soft modes. This
stochastic force is an important ingredient for the equilibration of
the classical modes.
The standard procedure \cite{21} is to introduce a stochastic
variable $\xi$ for each imaginary contribution to $S_{\rm IF}$, making 
use of the identity 
\begin{equation}
e^{-{1\over 2}\chi^TA\chi} = [{\rm det}(2\pi A)]^{-1/2} \int {\cal D}\xi
\; e^{-{1\over 2}\xi^TA^{-1}\xi + i\chi^T\xi}. \label{stovar}
\end{equation}
Here $A$ is a symmetric matrix, $\chi$ and $\xi$ are vectors, and we have
employed obvious vector notation.  Applying this technique here we
need to introduce three separate stochastic functions $\xi_N(x)$,
$N=1,2,3$, with the stochastic weights\footnote{We have chosen the
index of the stochastic field to match the power of the field
operators in the stochastic action.  In the notation of ref.
\cite{12}: $\xi_2\to\xi_1,\; \xi_1\to\xi_2$; in that of ref. \cite{18}.
$\xi_1\to -\eta,\; \xi_2\to -\xi,\; \xi_3\to-\nu$.}
\begin{equation}
P[\xi_N] = C_N \exp\left\{ \int d^4xd^4y\; \xi_N(x) \left[
-{(2N)!\over 4N^2}\; \left( {\rm Im} [\Gamma_{2N}^{k_c}]\right)^{-1}_{x,y}
\right] \xi_N(y)\right\}, \label{stowei}
\end{equation}
where $C_N$ are appropriate normalization constants.  The influence
functional then takes the form
\begin{equation}
e^{iS_{\rm IF}[\phi,\phi']} = \int {\cal D}\xi_1 {\cal D}\xi_2 {\cal
D}\xi_3\; P[\xi_1]P[\xi_2]P[\xi_3]\; e^{i\tilde S_{\rm
IF}[\phi,\phi';\xi_i]} \label{inffunc}
\end{equation}
with the new stochastic influence action
\begin{equation}
\tilde S_{\rm IF}[\phi,\phi';\xi_1,\xi_2,\xi_3] = {\rm Re}\left( S_{\rm IF}
[\phi,\phi']\right) + \sum_{N=1}^3 {1\over N} \int_{t_i}^{t} d^4x
\; R_N^-(x) \xi_N(x). \label{infact}
\end{equation}
This expression forms the basis for the classical equations of motion
for the soft modes $\phi(x)$.

\subsection{Classical Equations of Motion}

As we argued in the introduction, the dynamics of the
soft modes under the stochastic action
\begin{equation}
S[\phi] - S[\phi'] + \tilde S_{\rm IF}[\phi,\phi';\xi] \label{x}
\end{equation}
is governed by (quasi-)classical physics, if we assume that the soft modes
are sufficiently populated. As we will argue later this assumption will be 
justified by the full dynamics which we will derive and discuss
in the following. In addition, as sketched briefly in appendix A, we note 
that possible initial quantum mechanical correlations in the system will 
decohere due to the stochastic forces acting randomly on the soft modes and 
so destroying those correlations (or phases) with time.  We will not address
this issue here, as in the present case we are convinced that the soft modes 
behave already classically, and thus refer the reader to the extensive
literature on decoherence \cite{10,18,Zu82,Hal89,Alb92,GMH93}.
Loop contributions to the effective action $\Gamma_{\rm CTP}[\bar\phi]$ 
from the soft modes (under the influence of $\tilde S_{\rm IF}$) are not 
necessarily negligible, but at high temperature they are dominated by 
thermal fluctuations, not quantal ones, which are relatively suppressed 
by a factor $\hbar k_c/T$.  Thermal loops correspond to tree diagrams 
with thermally populated external legs.  They can, therefore, be accounted 
for by averaging the classical equations of motion at the tree level over 
a thermal ensemble of initial states.

The classical equations of motion for the soft modes are obtained by
a generalized stationary phase approximation (see appendix A) for the 
effective action defined on the time contour path ${\cal C}$ by first 
varying (\ref{x}) with respect to $\phi_{\Delta}\equiv (\phi-\phi')$ and 
then setting $\phi=\phi'$.

Making use of the
relations
\begin{eqnarray}
\left . {\delta R_N^-(x) \over \delta\phi_{\Delta}(y)} 
\right\vert_{\phi=\phi'} &= &N\phi(x)^{N-1} \delta(x-y), \nonumber \\
\left . {\delta R_N^+(x) \over \delta\phi_{\Delta}(y)}
\right\vert_{\phi=\phi'} &= &0, \qquad 
\left . R_N^+(x) \right\vert_{\phi=\phi'} = 2\phi (x)^N, \label{use}
\end{eqnarray}
we find:
\begin{eqnarray}
\sqcap{\mkern-12mu \sqcup} \phi(x) &+ &\tilde m^2 \phi(x) +
{\tilde g^2\over 6} \phi^3(x) + \sum_{N=1}^3 {1\over (2N-1)!} \phi(x)^{N-1}
\int_{t_i}^{t} d^4y\; {\rm Re}\left[ \tilde\Gamma_{2N}^{k_c}(x-y)\right]
\phi(y)^N \nonumber \\
&= &\sum_{N=1}^3 \phi(x)^{N-1} \xi_N(x)\, \, \, . \label{EOM}
\end{eqnarray}

We first note that the resulting equation of motion for the soft
fields is explicitly causal as the real part of the influence action
(\ref{expanded}) contains a step function assuring causality, i.e.
\begin{equation}
\theta (x_0-y_0) {\rm Re} \left[ \Gamma_{2N}^{k_c}(x-y)\right]
\end{equation}
defines just the response function (to lowest non-vanishing order)
of the hard modes on the soft modes and thus corresponds to the familiar
Kubo relation of systems near equilibrium generalized to the present case.

Secondly, the real part of the influence action is divergent and must be
renormalized.  Since the ultraviolet divergences of the vertex function 
$\Gamma_{2N}^{k_c}$ are independent of the low-momentum cut-off
$k_c$ and of the temperature, they can be absorbed in the standard
counter terms for mass and coupling constant renormalization.  We can,
therefore, replace the divergent $\Gamma_{2N}^{k_c}(x-y)$ by the
regularized expression 
\begin{equation}
\tilde\Gamma_{2N}^{k_c}(x-y) = \Gamma_{2N}^{k_c}(x-y) - G_{2N} \delta(x-y)
\label{regular}
\end{equation}
where $G_{2N}$ subtracts the temperature independent short-distance
singularity of $\Gamma_{2N}^{k_c=0}$.  In our specific case here,
$\Gamma_2^{k_c}$ and $\Gamma_4^{k_c}$ require renormalization.  
$\Gamma_6^{k_c}$ does not because it is finite and even vanishes in the 
limit $k_{c}\to 0$.  Hence, $\tilde m$ and $\tilde g$ denote the renormalized
mass and coupling constant, respectively, and $\tilde\Gamma_{2N}^{k_c}$ 
are the renormalized vertex functions.  The tadpole contributions of Figure 
1(a) and 1(b) to $\tilde\Gamma_2$ contribute thermal terms to the effective 
mass,
changing $\tilde m$ to $m^*$.  The thermal effective mass $m^*$ will also 
depend on the momentum cutoff $k_c$ as only the mean field contributions of 
the hard modes are accounted for.  Equation (\ref{EOM}) is graphically
represented in Figure 2.  The terms on the right-hand side represent
stochastic source, mass, etc., terms for the soft field modes.  In
particular, the term $\xi_1(x)$ indicates that the soft modes do not
obey a source-free equation, rather, they are in constant contact with
randomly fluctuating sources generated by the hard modes that have
been integrated out.

The equation of motion (\ref{EOM}) resembles a Langevin equation 
analogous to the situation of quantum Brownian motion.  However, 
as it stands, it is clearly nonlocal in time,  and one has to identify
the appropriate Markovian limit (or approximation) of this equation.
Indeed, the question arises whether a Markovian limit exists at all.
The answer to this question will have a bearing on the detailed 
understanding of dissipation in quantum field theories.  Clearly, if a
Markovian limit exists, it will provide a much simpler equation
which can be applied for practical purposes.

\section{Dissipation}

\subsection{Origin of Dissipation}

In this section we elucidate the dissipative character of the resulting 
contributions to the semiclassical equation of motion (\ref{EOM}) 
deriving from the real parts of the influence action at order
$\tilde{g}^4$.
Only the contributions of ${\rm Re}\left( S_{\rm IF}^{(c)} \right)$,
${\rm Re}\left( S_{\rm IF}^{(d)} \right)$ and
${\rm Re}\left( S_{\rm IF}^{(e)} \right)$ will be responsible for
dissipation, whereas the other contributions only renormalize
the effective mass parameter up to order $\tilde{g}^4$. With this in mind
we concentrate on their influence on the equation of motion (\ref{EOM})
of the soft modes $\phi (x)$:
\begin{eqnarray}
\sqcap{\mkern-12mu \sqcup} \phi(x) &+ &m^{*\, 2}\phi(x) +
{\tilde g^2\over 6} \phi^3(x)
+ \left( \sum_{i=c,d,e} (-1) \,
\left. \frac{\delta {\rm Re} \left( S_{\rm IF}^{(i)}[\phi , \phi_{\Delta }]
\right)}{\delta \phi_{\Delta }(x)} \right| _{\phi_{\Delta }=0} \right)
\nonumber \\
&= & \left( \xi_1(x) \, + \, \phi (x) \xi_2(x) \, + \,
\phi (x)^2 \xi_3(x) \right)
\, \, \, . \label{EOMa}
\end{eqnarray}
According to the general fluctuation-dissipation theorem, the three terms 
on the left-hand side lead to the dissipation of energy stored in the soft 
modes (the ``system'') to the hard modes (the ``bath'') balancing the 
energy gained continuously from the fluctuating force terms on the 
right-hand side, which were obtained from the imaginary parts of the 
influence functional  ${\rm Im}\left( S_{\rm IF}^{(c)} \right)$,
${\rm Im}\left( S_{\rm IF}^{(d)} \right)$, and
${\rm Im}\left( S_{\rm IF}^{(e)} \right)$. 

 From Appendix (B) we immediately find for the three dissipative terms
in the sum of eq. (\ref{EOMa}):
\begin{eqnarray}
- \left. \frac{\delta {\rm Re} \left( S_{\rm IF}^{(c)} \right)}
{\delta \phi_{\Delta }(x)} \right| _{\phi_{\Delta }=0}
&=& \int_0^{\infty} d\tau \int d^3\tilde{x} \,
\left(\frac{1}{6} \tilde{g}^4 \right)
\left( G_>^{k_c} (\tilde{{\bf x}},\tau)^3 -
G_<^{k_c}(\tilde{{\bf x}},\tau)^3 \right) \,
\phi ({\bf x}-\tilde{{\bf x}},t-\tau ) \;,
\nonumber \\
- \left. \frac{\delta {\rm Re} \left( S_{\rm IF}^{(d)} \right)}
{\delta \phi_{\Delta }(x)} \right| _{\phi_{\Delta }=0}
&=& \phi (x) 
\int_0^{\infty} d\tau \int d^3\tilde{x} \,
(\frac{i}{4} \tilde{g}^4) \left( G_>^{k_c} (\tilde{{\bf x}},\tau)^2 -
G_<^{k_c}(\tilde{{\bf x}},\tau)^2 \right) \,
\phi ({\bf x}-\tilde{{\bf x}},t-\tau )^2  \;,
\label{termscde} \\
- \left. \frac{\delta {\rm Re} \left( S_{\rm IF}^{(e)} \right)}
{\delta \phi_{\Delta }(x)} \right| _{\phi_{\Delta }=0}
&=& \phi (x)^2 \int_0^{\infty} d\tau \int d^3\tilde{x} \,
(-\frac{1}{12} \tilde{g}^4) \left( G_>^{k_c} (\tilde{{\bf x}},\tau) -
G_<^{k_c}(\tilde{{\bf x}},\tau) \right) \,
\phi ({\bf x}-\tilde{{\bf x}},t-\tau )^3 \;, \nonumber
\end{eqnarray}
where we have set $x_0=t$.

Obviously, the contributions of the hard modes have a rather simple structure:
The integration kernels are causally weighting the difference between powers
of the real-time propagators $G_>^{k_c}$ and $G_<^{k_c}$.  For convenience
we define the {\em memory} kernels ${\cal M}^{(i)}$:
\begin{eqnarray}
{\cal M}^{(c)}(\tilde{{\bf x}},\tau ) &=&
\frac{1}{6} \tilde{g}^4 \left( G_>^{k_c} (\tilde{{\bf x}},\tau)^3 -
G_<^{k_c}(\tilde{{\bf x}},\tau)^3 \right) \, \, \, ,
\nonumber \\
{\cal M}^{(d)}(\tilde{{\bf x}},\tau ) &=&
\frac{i}{4} \tilde{g}^4 \left( G_>^{k_c} (\tilde{{\bf x}},\tau)^2 -
G_<^{k_c}(\tilde{{\bf x}},\tau)^2 \right) \, \, \, ,
\label{memcde} \\
{\cal M}^{(e)}(\tilde{{\bf x}},\tau ) &=&
-\frac{1}{12} \tilde{g}^4 \left( G_>^{k_c} (\tilde{{\bf x}},\tau) -
G_<^{k_c}(\tilde{{\bf x}},\tau) \right) \, \, \, . \nonumber
\end{eqnarray}
To further evaluate the dissipative terms (\ref{termscde}) one has to 
insert an approximation for the time dependence of the soft modes
$\phi ({\bf x}-\tilde{\bf x},t-\tau)$ under the integrals.  Previous
authors \cite{9,12,Mo90,MS85} have expanded $\phi$ around $t$ up to 
the first order gradient correction, i.e.:
\begin{equation}
\phi ({\bf x}- \tilde{{\bf x}},t-\tau)^n \, \approx \,
\phi ({\bf x},t)^n - n \tau\phi ({\bf x},t)^{n-1} \dot{\phi }({\bf x},t)
\, \, \, . \label{inst}
\end{equation}
This amounts to a quasi instantaneous approximation of the soft modes in the
Heisenberg picture.  If this approximation is made, the dissipative
terms (\ref{termscde}) vanish, unless an explicit width $\Gamma$ is
introduced\footnote{The inconsistency of this approach becomes apparent
when one compares the memory time of the dissipative kernels
with the characteristic oscillation period of the soft modes
$((m^{*2}+k_c^2)^{-1/2})$. From (\ref{memcde})
the extension in past time of the kernels can roughly
be estimated to be of the same order, i.e. $k_c^{-1}$, as it sets the
the typical scale. (The memory time
in a scattering process is given by the uncertainty principle as
$\tau_{\rm mem} \approx 1/\langle \Delta E \rangle $, where
$\langle \Delta E \rangle$ denotes the averaged energy transferred in
the reaction. For the soft modes the energy transferred cannot be larger
than $k_c$.) Hence the soft modes will oscillate entirely over the
extension of the memory kernels.}  in the propagators $G_{>,<}^{k_c}$ 
of the hard modes \cite{9,12}.

We believe that this procedure obscures the true dissipative character 
of the memory terms.  In fact, as we will see below, the Fourier 
frequencies of the soft modes are essential parts of the correct strategy 
for revealing the close connection of the expressions (\ref{termscde}) 
with the more familiar on-shell scattering rates of soft modes on hard 
particles. Only if the soft modes oscillate periodically with some given 
frequency, can they scatter on the hard momentum modes or, if the 
frequencies allow for this, even produce hard particles.  In a certain
sense, the soft modes need time to scatter and again come on shell.
The instantaneous approximation (\ref{inst}) clearly does not account
for the oscillatory behavior of the soft modes and hence cannot properly
establish the correct Markovian limit of the expressions (\ref{termscde}).
The main objective of the remainder of this section is the derivation of
a well-defined Markovian form of the dissipative terms (\ref{termscde})
that takes the oscillatory character of the soft modes in account.

The soft modes will, of course, oscillate with a certain spectrum of
frequencies. This is true also for the spatial Fourier modes 
$\phi ({\bf k},t)$.  In a linear approximation one expects, however, 
that they oscillate with some pronounced frequency
\begin{equation}
E_k \, \approx \, \sqrt{m^{*2}+ {\bf k}^2}\;\; . \label{specfunc}
\end{equation}
This should be rigorously valid in the weak coupling regime when the
soft modes propagate almost freely. Without further specifying the 
single-particle energy $E_k$, we can approximately express the soft modes 
at earlier times $(t-\tau )$ by their value at the time $t$ in the linear 
harmonic approximation
\begin{equation}
\phi ({\bf k},t-\tau ) \approx  \phi ({\bf k},t) \cos E_k\tau \, - \,
\dot{\phi }({\bf k},t) \frac{1}{E_k}\sin E_k\tau \;\;. \label{lha}
\end{equation}
Note that the $\dot{\phi }({\bf k},t)$ contribution turns out to be 
essential and cannot be neglected. Comparing (\ref{lha}) with (\ref{inst}) 
one immediately recognizes that the harmonic approximation (\ref{lha}) 
contains much more information about the past than the instantaneous 
approximation (\ref{inst}). Equation (\ref{lha}) can be considered as
a Markov approximation to the soft field modes in the interaction 
picture, where the free oscillation frequency has been extracted.

To reveal the full meaning of the three dissipative terms (\ref{termscde}) 
in the equation of motion (\ref{EOMa}) of the soft modes we insert the 
harmonic approximation (\ref{lha}) and further evaluate the three terms 
approximately.  Our strategy is completely equivalent to the quasi-particle 
approximation used in the more familiar kinetic theories when evaluating 
the collision term in transport processes \cite{DeG}.  A certain mode 
(quasi-particle) is assumed to propagate with a specific frequency, the 
quasi-particle energy.  This approximation is valid if the spectral function
of the quasi-particle is sharply peaked at this frequency. Moreover, 
pursuing the analogy with standard kinetic theory, the equation of motion 
(\ref{EOMa}) becomes instantaneous as all soft quasi-particle modes 
$\phi({\bf k},t)$ are evaluated at the same time.  We thus obtain the 
Markovian limit of eq. (\ref{EOMa}) when evaluating the time integral
over the memory kernels defined in (\ref{memcde}) with the soft modes 
propagating according to (\ref{lha}).

We write (\ref{termscde}) as
\begin{eqnarray}
- \left. \frac{\delta {\rm Re} \left( S_{\rm IF}^{(c)} \right)}
{\delta \phi_{\Delta }(x)} \right| _{\phi_{\Delta }=0}
&=&
\int \frac{d^3k}{(2\pi )^3} \, e^{i{\bf k}\cdot {\bf x}} \,
\Upsilon ^{(c)}({\bf k},t) \;, \nonumber \\
- \left. \frac{\delta {\rm Re} \left( S_{\rm IF}^{(d)} \right)}
{\delta \phi_{\Delta }(x)} \right| _{\phi_{\Delta }=0}
&=& \phi (x) \int \frac{d^3k}{(2\pi )^3} \, e^{i{\bf k}\cdot {\bf x}} \,
\Upsilon ^{(d)}({\bf k},t) \;, \label{Rewcde} \\
- \left. \frac{\delta {\rm Re} \left( S_{\rm IF}^{(e)} \right)}
{\delta \phi_{\Delta }(x)} \right| _{\phi_{\Delta }=0}
&=& \phi (x)^2 \int \frac{d^3k}{(2\pi )^3} \, e^{i{\bf k} \cdot {\bf x}} \,
\Upsilon ^{(e)}({\bf k},t) \; , \nonumber
\end{eqnarray}
with the notations
\begin{eqnarray}
\Upsilon ^{(c)}({\bf k},t) &=&
\theta (k_c-|{\bf k}|)
\int_0^{\infty } d\tau \, \phi ({\bf k},t-\tau ) \, {\cal M}^{(c)}({\bf k},\tau 
)
\;, \nonumber \\
\Upsilon ^{(d)}({\bf k},t) &=&
\int^{k_c} \frac{d^3k_1}{(2\pi )^3}
\theta (k_c-|{\bf k}-{\bf k}_1|)
\int_0^{\infty } d\tau \, \phi ({\bf k}-{\bf k}_1,t-\tau )
\phi ({\bf k}_1,t-\tau )
\, {\cal M}^{(d)}({\bf k},\tau) \;, \label{upscde} \\
\Upsilon^{(e)}({\bf k},t) &=& \int^{k_c} \frac{d^3k_1d^3k_2}{(2\pi )^6}
\theta (k_c-|{\bf k}-{\bf k}_1-{\bf k}_2|)
\int_0^{\infty } d\tau \, \phi ({\bf k}-{\bf k}_1-{\bf k}_2,t-\tau )
\phi ({\bf k}_1,t-\tau ) \phi ({\bf k}_2,t-\tau ) \, 
{\cal M}^{(e)}({\bf k},\tau ) \;, \nonumber
\end{eqnarray}
where the  momentum integrations for all soft modes are restricted by 
the cut-off momentum ${\bf k}_c$.  We can now insert the approximate form 
(\ref{lha}) for the soft modes.  It is convenient to first perform a Fourier 
transformation of the memory kernels 
\begin{equation}
{\cal M}^{(i)}({\bf k}, \omega) = \int_{-\infty}^{\infty} dt\; 
e^{i\omega\tau} {\cal M}^{(i)}({\bf k},\tau)
\end{equation}
and insert the expressions for the hard propagators of Appendix B. One finds
\begin{eqnarray}
{\cal M}^{(c)}({\bf k},\omega ) \, = \,
-\frac{\pi }{24}\tilde{g}^4\;(i) &\int_{k_c}& \frac{d^3q_1d^3q_2}{(2\pi )^6}
\, \frac{1}{\omega_1} \frac{1}{\omega_2} \frac{1}{\omega_3}
\, \theta (|{\bf k}-{\bf q}_1-{\bf q}_2|-k_c)
\nonumber  \\
& \times \left\{ \right. &
\left[ (1+n_1)(1+n_2)(1+n_3) \, - \, n_1 n_2 n_3 \right]
\delta (\omega - \omega_1 - \omega_2 -\omega_3)   \nonumber \\
&+& \left[ (1+n_1)n_2(1+n_3) \, - \, n_1 (1+n_2) n_3 \right]
\delta (\omega - \omega_1 + \omega_2 -\omega_3)   \nonumber \\
&+& \left[ n_1(1+n_2)(1+n_3) \, - \, (1+n_1) n_2 n_3 \right]
\delta (\omega + \omega_1 - \omega_2 -\omega_3) \nonumber \\
&+& \left[ n_1 n_2 (1+n_3) \, - \, (1+n_1) (1+n_2) n_3 \right]
\delta (\omega + \omega_1 + \omega_2 -\omega_3)  \nonumber \\
&+& \left[ (1+n_1) (1+n_2) n_3 \, - \, n_1 n_2 (1+n_3) \right]
\delta (\omega - \omega_1 - \omega_2 +\omega_3)  \nonumber \\
&+& \left[ (1+n_1) n_2 n_3 \, - \, n_1 (1+n_2) (1+n_3) \right]
\delta (\omega - \omega_1 + \omega_2 +\omega_3)  \nonumber \\
&+& \left[ n_1 (1+n_2) n_3 \, - \, (1+n_1) n_2 (1+n_3) \right]
\delta (\omega + \omega_1 - \omega_2 +\omega_3)  \nonumber \\
&+& \left[ n_1 n_2 n_3 \, - \, (1+n_1) (1+n_2) (1+n_3) \right]
\left. \delta (\omega + \omega_1 + \omega_2 +\omega_3) \right\}
\label{memc}
\end{eqnarray}
where $\omega_1=\omega_{{\bf q}_1}, \omega_2=\omega_{{\bf q}_2},\omega_3 = 
\omega_{{\bf k}-{\bf q}_1-{\bf q}_2}$ and $n_{\alpha}= 
n(\omega_{{\bf q}_{\alpha}})$, and
\begin{eqnarray}
{\cal M}^{(d)}({\bf k},\omega )  \, = \, - \frac{\pi }{8}\tilde{g}^4 \; (i)
&\int_{k_c}& \frac{d^3q}{(2\pi )^3}\, \frac{1}{\omega _1} \frac{1}{\omega _2}
\, \theta (|{\bf k}-{\bf q}|-k_c)
\nonumber \\
&\times \left\{ \right. &
\left[ (1+n_1)(1+n_2) \, - \, n_1 n_2 \right]
\delta (\omega - \omega_1 - \omega_2 )   \nonumber \\
&+& \left[ n_1 (1+n_2) \, - \, (1+n_1) n_2 \right]
\delta (\omega + \omega_1 - \omega_2 )   \nonumber \\
&+& \left[ (1+n_1)n_2 \, - \, n_1 (1+ n_2) \right]
\delta (\omega - \omega_1 + \omega_2 )  \nonumber \\
&+& \left[ n_1 n_2  \, - \, (1+n_1) (1+n_2) \right]
\left. \delta (\omega + \omega_1 + \omega_2) \right\} \label{memd}
\end{eqnarray}
where $\omega_1 = \omega_{{\bf q}}, \;\omega_2 = \omega_{{\bf k}-{\bf q}}$,
and
\begin{eqnarray}
{\cal M}^{(e)}({\bf k},\omega )&= &- \frac{\pi }{12}\tilde{g}^4 \, (i)
\frac{1}{\omega_{\bf k}}\; \theta(\vert{\bf k}\vert - k_c) \cdot \nonumber \\
&&\left\{ \left[ (1+n(\omega_{\bf k})) - n(\omega_{\bf k}) \right]
\delta (\omega - \omega_{\bf k}) + \left[ n(\omega_{\bf k}) - 
(1+n(\omega_{\bf k}))  \right] \delta (\omega + \omega_{\bf k}) \right\}. 
\label{meme}
\end{eqnarray}
The memory kernels ${\cal M}^{(i)}({\bf k},\omega )$ have a very simple
structure. They all are purely imaginary and satisfy the relation
\begin{equation}
{\cal M}^{(i)}({\bf k},\omega )\, = \,
-{\cal M}^{(i)}({\bf k},-\omega )\; .  \label{prop}
\end{equation}
Moreover, the kernels are directly related to the discontinuities of the
respective self-energy insertions of the hard modes.  For a detailed 
discussion and interpretation of such discontinuities in finite temperature
field theory we refer the reader to the article by Weldon \cite{We83}.
 From the expressions (\ref{memc}-\ref{meme}) we see that each
kernel is composed of two contributions: the direct part $\Gamma _{d}({\bf k},
\omega )$, or the ``loss'' term, and the inverse part $\Gamma _{i}({\bf k},
\omega )$, or ``gain'' term.  Following \cite{We83}, we write
\begin{equation}
i{\cal M}({\bf k},\omega) \propto \Gamma _{d}({\bf k},\omega ) -
\Gamma _{i}({\bf k},\omega)\; , \label{We1}
\end{equation}
where $\Gamma _d$ and $\Gamma _i$, respectively, are positive and can be
identified with the decay rate of a quasi-particle with momentum ${\bf k}$ 
and energy $\omega$ into the allowed open channels or the production rate
due to the inverse processes.  In addition, the detailed balance relation 
\begin{equation}
\frac{\Gamma_d(\omega )}{\Gamma_i(\omega )}  =  \exp (\omega /T)
\label{We2}
\end{equation}
follows immediately if the distribution function $n(\omega )$ of the hard
modes is assumed to be in thermal equilibrium.

This identification represents an important step toward understanding
the dissipative nature of the terms (\ref{termscde}).  Intuitively, we
expect dissipation to be closely related to the on-shell scattering rates 
of the soft modes when interacting with the hard thermal particles.  We 
have to manipulate the terms further in order to fully expose this 
connection.

We now insert the approximation (\ref{lha}) for the soft quasi-particle 
modes into the semi-infinite integrals $\Upsilon ^{(i)} ({\bf k},t)$ given
by (\ref{upscde}). In order to evaluate the time integral we express the 
memory kernels by their Fourier transform ${\cal M}^{(i)}({\bf k},\omega )$
and make use of the relation
\begin{equation}
\int_0^{\infty } d\tau \,
e^{i(\gamma-\omega)\tau} = i \lim_{\epsilon \rightarrow 0^+} 
\frac{1}{(\gamma - \omega )+i\epsilon }
= i {\it P} \frac{1}{\gamma - \omega } + \pi \delta (\gamma - \omega )
\; .  \label{dirrel}
\end{equation}
The remainder of the calculation is straightforward but cumbersome.
One has to express the sine and cosine contributions by exponentials
and collect all terms. Relation (\ref{prop}) helps in identifying
certain pairwise identical contributions. The final result reads
\begin{eqnarray}
\Upsilon^{(c)} ({\bf k},t) &=&
\phi({\bf k},t) \,
\theta (k_c-|{\bf k}|)
\int \frac{d\omega }{2\pi } \; i
{\cal M}^{(c)}({\bf k}, \omega) \, {\it P}\frac{1}{E_k - \omega}
+ \dot{\phi } ({\bf k},t) \,
\theta (k_c-|{\bf k}|)
\frac{1 }{2E_k } \,
i {\cal M}^{(c)}({\bf k}, E_k)  \, \, \, , \label{upsc} \\
\Upsilon ^{(d)} ({\bf k},t) &=&
\int^{k_c} \frac{d^3k_1}{(2\pi )^3} \,
\theta (k_c-|{\bf k}-{\bf k}_1|)
\left\{
\phi ({\bf k}-{\bf k}_1,t) \phi ({\bf k}_1,t) \frac{1}{2}
\;\int \frac{d\omega }{2\pi } \, i
{\cal M}^{(d)}({\bf k}, \omega) \right. \nonumber \\
&&\qquad\qquad \times \left( {\it P}\frac{1}{E_{{\bf k}-{\bf k}_1}+E_{k_1} -
\omega} 
+ {\it P}\frac{1}{E_{{\bf k}-{\bf k}_1}-E_{k_1} - \omega} \right)
\nonumber \\
&& \qquad  + \dot{\phi} ({\bf k}-{\bf k}_1,t) \dot{\phi } ({\bf k}_1,t)
\frac{1}{2 E_{{\bf k}-{\bf k}_1} E_{k_1}} \;\int\; 
\frac{d\omega }{2\pi } \, i
{\cal M}^{(d)}({\bf k}, \omega) \nonumber \\
&&\qquad\qquad \times \left. \left( -{\it P}\frac{1}{E_{{\bf k}-{\bf
k}_1}+E_{k_1} - 
\omega} + {\it P}\frac{1}{E_{{\bf k}-{\bf k}_1}-E_{k_1} - \omega} \right) 
\right\} \nonumber \\
&& + \int^{k_c} \frac{d^3k_1}{(2\pi )^3} \,
\dot{\phi } ({\bf k}-{\bf k}_1,t) \phi ({\bf k}_1,t)
\frac{1}{4 E_{{\bf k}-{\bf k}_1} }
\theta (k_c-|{\bf k}-{\bf k}_1|)
\nonumber \\
&&\qquad\qquad  \times \left(
i{\cal M}^{(d)}({\bf k},E_{{\bf k}-{\bf k}_1}+E_{k_1}) +
i{\cal M}^{(d)}({\bf k},E_{{\bf k}-{\bf k}_1}-E_{k_1}) \right) \, \, \, ,
\label{upsd} \\
\Upsilon ^{(e)} ({\bf k},t) &=&
\int^{k_c} \frac{d^3k_1d^3k_2}{(2\pi )^6} \,
\theta (k_c-|{\bf k}_3|)
\left\{ \phi ({\bf k}_1,t) \phi ({\bf k}_2,t) \phi ({\bf k}_3,t) \, 
\frac{1}{4} \; \int \frac{d\omega }{2\pi } \, (i)
{\cal M}^{(e)}({\bf k}, \omega) \right. \nonumber \\
&&\qquad  \times \left( {\it P}\frac{1}{E_1+E_2+E_3 - \omega} +
{\it P}\frac{1}{E_1+E_2-E_3 - \omega} +
2{\it P}\frac{1}{E_1-E_2+E_3 - \omega} \right)
\nonumber \\
&& + \quad \dot{\phi } ({\bf k}_1,t) \dot{\phi }({\bf k}_2,t)
\phi ({\bf k}_3,t) \, \frac{3}{4E_1E_2} \int \frac{d\omega}{2\pi} \;i\;
{\cal M}^{(e)}({\bf k}, \omega) \nonumber \\
&&\qquad \times \left. \left(- {\it P}\frac{1}{E_1+E_2+E_3-\omega}
-{\it P}\frac{1}{E_1+E_2-E_3 - \omega} + 
2{\it P}\frac{1}{E_1-E_2+E_3 - \omega} \right) \, \right\}
\nonumber \\
&&+ \int^{k_c} \frac{d^3k_1d^3k_2}{(2\pi )^6} \,
\theta (k_c-|{\bf k}_3|)
\left\{
\phi ({\bf k}_1,t) \phi ({\bf k}_2,t)
\dot{\phi }({\bf k}_3,t) \, \frac{3}{8E_3}  \right. \nonumber \\
&&\quad \times \left(
i{\cal M}^{(e)}({\bf k},E_1+E_2+E_3) + 
i{\cal M}^{(e)}({\bf k},-E_1-E_2+E_3) + 2i{\cal M}^{(e)}
({\bf k},E_1-E_2+E_3) \right)  \nonumber \\
&&\quad + \dot{\phi } ({\bf k}_1,t) \dot{\phi }({\bf k}_2,t)
\dot{\phi }({\bf k}_3,t) \, \frac{1}{8E_1 E_2 E_3} \nonumber \\
&&\qquad \times \left(
i{\cal M}^{(e)}({\bf k},-E_1-E_2-E_3) +
i{\cal M}^{(e)}({\bf k},E_1+E_2-E_3) \right. \nonumber \\
&&\qquad \quad \left. \left. +2i{\cal M}^{(e)}({\bf k},E_1-E_2+E_3) \right)
\right\}
. \label{upse} 
\end{eqnarray}
where ${\bf k}_3 \,= {\bf k} -{\bf k}_1 - {\bf k}_2$.

One notices that the contributions with an odd power in $\dot{\phi }$
contain the on-shell reaction rates (\ref{memc}-\ref{meme}).  Thus we 
expect them to describe the dissipative aspects of the equation of motion 
for the soft modes as energy is transferred by scattering (or production) 
processes between the soft and hard modes.  Since this interpretation is
most obvious for the first term $\Upsilon^{(c)}$, let us start by
discussing this term in more detail.  Fourier transforming the equation 
of motion (\ref{EOMa}) for the mode ${\bf k}$, we find from (\ref{Rewcde})
that ${\rm Re}(S^{(c)}_{{\rm IF}})$ contributes on the right hand side as
\begin{eqnarray}
&&\ddot\phi ({\bf k},t)+ {\bf k}^2\phi ({\bf k},t) + \tilde{m}^2 \phi 
({\bf k},t) + \frac{\tilde{g}^2}{6} \int^{k_c} {d^3k_1d^3k_2\over (2\pi)^6}
\theta (k_c-|{\bf k}-{\bf k}_1-{\bf k}_2|)
\phi({\bf k}_1,t)\phi({\bf k}_2,t) \phi({\bf k}-{\bf k}_1-{\bf k}_2,t)
\nonumber \\
&&\quad +\phi({\bf k},t) P \int \frac{d\omega }{2\pi }\; 
\frac{i{\cal M}^{(c)}({\bf k}, \omega)}{E_k-\omega}   + 
\dot{\phi} ({\bf k},t) \, \frac{i {\cal M}^{(c)}({\bf k},E_k)}
{2E_k} +  \ldots \, = \, \xi_1({\bf k},t) + \ldots\; . \label{EOMb}
\end{eqnarray}
The first additional term can be readily interpreted as momentum 
dependent contribution to the mass term (after proper renormalization).
The second term has the familiar structure of a velocity dependent damping 
term, i.e. $\eta \dot{\phi }({\bf k},t)$.  From (\ref{memc}) and the
detailed balance relation (\ref{We2}) we find that the friction coefficient
\begin{equation}
\eta ^{(c)}({\bf k}) \, = \, \frac{i{\cal M}^{(c)}({\bf k},E_k)}{2E_k}
\ge 0 \label{dampc}
\end{equation}
is in fact positive.  It is exactly related to the rate $(\Gamma_d - 
\Gamma_i)$ with which a small disturbance relaxes towards equilibrium 
by coupling to the various open channels \cite{We83}. The damping rate
for the soft modes in this particular channel follows immediately as
\begin{equation}
\gamma ^{(c)}({\bf k}) \, = \, \frac{1}{2}
\eta ^{(c)}({\bf k}) \, \, \, ,
\label{dampratec}
\end{equation}
as the left hand side of eq. (\ref{EOMb}) has the well known solution
of a weakly damped oscillator
$$
\phi ({\bf k},t) \, \sim \, e^{-\eta^{(c)}({\bf k})t/2} \cos
(E_k^2 - \frac{1}{4}\eta^{(c)}({\bf k})^2)^{1/2}t \, \, \, .
$$

Generalizing this observation we expect a similar interpretation
of the two other dissipative terms, $\Upsilon^{(d)}$ and $\Upsilon^{(e)}$,
to hold: (1) The principal value contributions give rise to a (generally 
momentum dependent) vertex renormalization and to a new (again momentum 
dependent) six-point vertex;  (2) the on-shell contributions can be 
associated with damping processes involving two hard particles and two soft 
modes or involving one hard particle interacting with three soft modes, 
respectively.  In Appendix C we present a further analysis of the 
contributions from ${\rm Re}(S^{(d)}_{{\rm IF}})$ and 
${\rm Re}(S^{(e)}_{{\rm IF}})$, where a bridge is cast to the evaluation
of the discontinuities by means of soft quasi-particle propagators rather 
than soft fields. In particular the amplitude squared is easily identified 
with the quasi-particle density of the soft modes, i.e.
\begin{equation}
|\phi ({\bf k},t)|^2  = 
\phi ({\bf k},t) \phi (-{\bf k},t)  \approx 
\frac{1}{E_k} \, N({\bf k},t)  \approx 
\frac{1}{E_k} \, (1+N({\bf k},t)) \label{ident}
\end{equation}
in the quasi-classical regime where the soft modes are strongly populated.
One finds:
\begin{eqnarray}
\eta ^{(d)}({\bf k},t) & \approx & \frac{1}{4E_k}
\int^{k_c} \frac{d^3k_1}{(2\pi )^3} \,
|\phi ({\bf k}_1,t)|^2 \nonumber \\
&&\quad \times \left(
i{\cal M}^{(d)}({\bf k}+{\bf k}_1,E_k+E_{k_1}) +
i{\cal M}^{(d)}({\bf k}+{\bf k}_1,E_k-E_{k_1}) \right) \;, \label{dampd}  \\
\eta ^{(e)}({\bf k},t) &\approx &\frac{3}{4E_k}
\int^{k_c} \frac{d^3k_1d^3k_2}{(2\pi )^6}\,
|\phi ({\bf k}_1,t)|^2 \; |\phi ({\bf k}_2,t)|^2  \nonumber \\
&&\quad \times \left(
i{\cal M}^{(e)}({\bf k}+{\bf k}_1+{\bf k}_2,E_k+E_{k_1}+E_{k_2}) \right.
\nonumber \\
&&\qquad \qquad  + i{\cal M}^{(e)}({\bf k}+{\bf k}_1+{\bf k}_2,E_k-E_{k_1}
-E_{k_2}) \nonumber \\
&& \qquad \qquad \left.
+2i{\cal M}^{(e)}({\bf k}+{\bf k}_1+{\bf k}_2,E_k+E_{k_1}-E_{k_2})
\right) \, \, \, . \label{dampe}
\end{eqnarray}
These are the obvious generalizations of (\ref{dampc}) and correspond
to the friction coefficients for the possible on-shell scattering
or production processes
among two soft and two hard modes and three soft modes and one hard mode,
respectively.  We note, however, that these coefficients do not need to be 
positive for all times.  From (\ref{We1}) and (\ref{We2}) it follows that 
${\cal M}$ is positive for $\omega >0$, whereas ${\cal M} = - 
{\cal M}(|\omega |)$ becomes negative for $\omega <0$.  Hence the second 
integrand of (\ref{dampd}) as well as the last two of (\ref{dampe}) can 
become negative, so that in principle ``anti-damping'' can occur in some 
cases. This is not a conceptual difficulty.  Suppose one would prepare an 
initial system where a few modes are much less populated than all the 
others.  These special modes should grow in magnitude by scattering 
processes either due to the direct $\phi^3$-term in (\ref{EOMa}), or by 
scattering processes including other soft modes and hard particles. Energy 
is then transferred from the other soft modes to those few and into the 
reservoir of hard particles.

Summarizing this section, we succeeded in properly identifying the origin 
of dissipation from on-shell scattering processes of soft quasi-particles.
Dissipation corresponds directly to the imaginary parts of the
self energy insertions $\left[{\rm Re}\Gamma^{k_c}_{2N}\right]({\bf k},
\omega )$ in the response term of eq. (\ref{EOM})---which should not be 
mixed up with the imaginary parts of (\ref{expanded}) of the effective 
action---as one would have expected from standard finite temperature field 
theory.  Furthermore instantaneous, but momentum dependent dissipation 
coefficients have been derived in the proper Markovian limit of the 
effective terms (\ref{termscde}).

\subsection{The limit $k_c \rightarrow 0$}

To draw comparison with other work \cite{9,12,14} it is useful
to consider the limit $k_c\rightarrow 0$ for the scalar and massive theory:
\begin{equation}
\phi ({\bf x},t) \longrightarrow \phi (t); \quad
\phi({\bf k},t)\longrightarrow \phi (t) (2\pi )^3\delta^3({\bf k}) \;.
\label{kclim}
\end{equation}
The analysis follows in complete analogy to the preceding section so that 
we skip the intermediate steps and state the final results $(k_c\to 0)$:
\begin{eqnarray}
\eta^{(c)} & \longrightarrow & \frac{1}{2E_0}
i{\cal M}^{(c)}(0, E_0) \ge  0  \;, \label{damp0c} \\
\eta^{(d)}(t) &\longrightarrow &\phi (t)^2 \, \frac{1}{4E_0}
i{\cal M}^{(d)}(0,2E_0) \nonumber \\
&=& \phi (t)^2 \, \frac{\pi\tilde g^4}{32E_0}
\int_{k_c\rightarrow 0} \frac{d^3q}{(2\pi )^3} \,
\frac{1}{\omega _q^2} \, \left( 1+2n(\omega _q) \right)\,
\delta (E_0 - \omega _q) \, \, \, \ge \, 0  \, \, \, ,
\label{damp0d} \\
\eta^{(e)}(t) &\longrightarrow & 0   \, \, \, . \label{damp0e}
\end{eqnarray}
Here $E_0$ denotes the infrared limit of the dispersion relation $E_{\bf k}$
of the soft modes.  The first dissipation coefficient due to the ``sunset'' 
graph (Figure 1c) contains contributions due to scattering 
as well as particle production.  The second coefficient is only
nonvanishing when producing hard particles, i.e. when the low momentum 
mode $\phi (t)$ possesses Fourier frequencies that allow for such a 
dissipative channel. The last term vanishes as the hard propagator is 
restricted to momenta $|{\bf q}|>k_c=0$.

The dynamical mass $(m^*)^2$ to order $\tilde{g}^2$ is given by the
tadpole diagram (a) in Figure 1: $(m^*)^2=\tilde{g}^2T^2/24$,
which is the dominant contribution to the mass in the high temperature 
limit, i.e. when $\tilde{m} \ll m^*$.  Inserting this for
the dispersion relation of the energy, i.e. $E_0 = m^*$, the
damping coefficient $\eta ^{(c})$ is exactly related to twice
the damping rate of plasmons in $\Phi^4$ theory \cite{Par92,Je95,EH95}:
\begin{equation}
\eta^{(c)} \, = \, 2 \gamma (m^*,0) \, = \,
\frac{\tilde{g}^3T}{32\sqrt{24}\pi} \,\,\, ,
\label{dampplasmon}
\end{equation}
as one would have expected by our arguments given at the end
of the last section.

We admit that our approximate treatment of a single harmonic approximation 
(\ref{lha}) for the soft mode may fail when $\phi (t)$ contains a whole 
set of frequencies, e.g. if one thinks of unstable phase transitions where 
$\phi(t)$ acts as the relevant order parameter. In this case, however, one 
can work with the explicit temporally nonlocal effective terms which one 
can easily write down from the expressions of the memory kernels at 
${\bf k}=0$. Such a procedure was undertaken by Boyanovsky et al. \cite{14}.
At zero temperature the authors examined in detail the damping associated 
with the contribution of the `fish-graph' (d) in a more self-consistent 
treatment by resumming its influence on the evolution of the
(ensemble) averaged order
parameter $\langle \phi (t) \rangle $.  They found that the dissipation was 
strongest in the Goldstone sector when (nearly) massless particles can be 
produced. Such a reasoning follows also immediately from our result 
(\ref{damp0d}) as then $\eta^{(d)}$ becomes nonvanishing even for $k_c\to 0$.  
As outlined above, we share the authors' point of view that an instantaneous 
approximation as assumed in (\ref{inst}) is not valid and cannot correctly 
account for the dissipation contained in the nonlocal expressions, such as 
(\ref{termscde}).

\subsection{A remark on the fluctuation-dissipation relation}

The dissipative terms and the noise terms are related by fluctuation-dissipation
relations. From Appendices A and B we see that the memory kernels ${\cal M}
\equiv D_R$ (the response functions) and the noise kernels $\nu \equiv D_I$ 
are closely related (cf. their definitions (\ref{A21})
or (\ref{B11}, \ref{B12}, \ref{B13})). On general
grounds intimate relations among these kernels exist, known as the generalized
fluctuation-dissipation relations (FDR). However, their explicit realizations
are quite complicated. For a detailed derivation which can be readily applied
to our case we refer to \cite{10}.
Because of the symmetry relations
${\cal M}({\bf k},\omega ) =-{\cal M}({\bf k},-\omega )$ and
$\nu ({\bf k},\omega ) = \nu ({\bf k},-\omega )$
one defines a new kernel $\gamma $ as
\begin{equation}
{\cal M}(\tau )= \frac{d}{d\tau } \gamma (\tau ) \; . \label{Hu1}
\end{equation}
Then a relation among $\gamma $ and $\nu $ can be obtained in the form
\begin{equation}
\nu (\tau) = \int_{-\infty}^{\infty} ds \,
K(\tau - s) \gamma (s) \; , \label{Hu2}
\end{equation}
which is just the generalized FDR.  In order to see how $K$
can be derived, we make use of (\ref{We2}) together with (\ref{Hu1}) to write
\begin{equation}
\gamma (\omega) = {i\over\omega}{\cal M}(\omega)  \propto
{1\over \omega} \left( \Gamma _{d}({\bf k},\omega) - \Gamma_{i}
({\bf k},\omega ) \right) \; . \label{Hu3}
\end{equation}
For the noise kernels a similar identification holds (compare the explicit 
forms given in Appendix B)
\begin{equation}
\nu (\omega) \, \propto  \,  \Gamma _{d}({\bf k},\omega ) +
\Gamma_{i}({\bf k},\omega)  \;. \label{Hu4}
\end{equation}
(The Fourier transform of the noise kernel is thus always non-negative.)
Because of (\ref{We2}), one finds that
\begin{equation}
K(\omega )  =  \frac{\gamma (\omega )}{\nu (\omega )} 
=  \frac{1}{\omega }\, 
\frac{\exp (\hbar \omega /T) -1}{\exp (\hbar \omega /T) +1} \; . \label{Hu5}
\end{equation}
Now the important point to stress is that the frequencies involved
for the soft modes are much smaller than the temperature T, i.e.
$E_k \ll T$ for $k_c \ll T$. Hence only those low frequencies
are relevant for the kernels ${\cal M}, \nu$ and $\gamma $, as also
only those noises containing small frequencies can influence the dynamics
of the soft modes. For $\nu \equiv I$ this follows in particular if one rewrites
the Gaussian distribution of the noise, e.g. (\ref{A29}), in the frequency
domain as
\begin{equation}
P[\xi(\omega )] = {1\over N} \exp \left( -{1\over 2} \int
d\omega
\xi(\omega ) I^{-1}(-\omega ) \xi(-\omega )\right). \label{A29freq}
\end{equation}
If one thus considers only the relevant low frequency part of the kernels,
$K$ becomes effectively constant and thus
local in time:
\begin{equation}
K(s) \, \stackrel{\omega \ll T}{\longrightarrow} \,2\, T/\hbar \; ,
\label{Hu6}
\end{equation}
the famous Kubo identity. Such a simple relation always occurs if the system
behaves classically \cite{Co85}. In fact, as we have argued before,
the soft modes should satisfy this condition.

The FDR ensures that the soft modes approach thermal equilibrium precisely
at the temperature $T$ of the hard modes, when they evolve under the
equation of motion (\ref{EOMa}).  The presence of these terms is
essential, as it forces the soft modes to thermalize at $T$ independent 
of their initial configuration. In particular, the noise terms will continuously
`heat' the system (for field theories see ref. \cite{CS91}) whereas the
dissipative part of the response function counteracts. Equilibrium is achieved
when the system has thermalized to the temperature dictated by the bath.
This aspect is not important if one only wants to study the time evolution 
of thermal configurations of the soft modes.  However, for the study of 
configurations far off thermal equilibrium the presence of both noise and 
friction terms is crucial, as it ensures that the soft modes will become 
populated thermally, i.e. their amplitudes become large. This justifies our
basic assumption.

This remark bears relevance to the issue whether thermal masses can
suppress chaos in certain nonlinear classical field theories as
suggested by Blaizot and Iancu \cite{BI94a}. They argued that small
amplitude, homogeneous field configurations in non-Abelian gauge
theories oscillate regularly due to their thermal mass, and only
large amplitude waves become chaotic. Our above result shows that
soft modes do not remain of small amplitude when stochastic forces
are included in their description, rather their amplitude always approaches
the thermal limit. It would be interesting to study this phenomenon,
e.g.~in the case of two massless scalar fields $\phi$ and $\chi$,
which are coupled through an interaction of the form $g^2\phi^2\chi^2$.
This theory exhibits chaos in the naive classical limit. Whether the
effective stochastic classical theory for soft field modes remains
chaotic in spite of thermal mass generation remains to be seen.

\subsection{Memory effects}

As discussed at length above, the real part of the influence action yields
nonlocal, dissipative terms in the effective classical equation of
motion (\ref{EOMa}). For further exploration of the nature of dissipation
we made the Markov ansatz (\ref{lha}) to calculate the dissipation 
coefficients. If the strength of the soft modes, $\sqrt{|\phi({\bf k},t)|^2}$,
also changes rapidly with time (e.g. in the strong coupling regime) the 
Markov limit may not be valid.  This can happen as the memory kernels 
${\cal M}(\tau )$ have support in the past roughly given by the scattering 
time of the individual processes.  To avoid such interference effects one 
has to choose the appropriate cutoff $k_c$ large enough to restrict the 
temporal extent of the kernels. This becomes relevant in the case of
massless theories in the strong coupling regime. However, the validity of 
the quasi-classical behavior of the soft modes is then no longer warranted.

\subsection{Infrared behavior}

We restate the quasi-classical equation of motion for the soft modes 
$\phi({\bf k},t)$ in their approximate instantaneous form:
\begin{eqnarray}
{\partial^2\phi\over\partial t^2} + && \left ( {\bf k}^2+ \tilde m^2 
+\sum_{i=a,b,c} \mu_{1,k_c}^i \right)\phi + \left(
{\tilde g^2\over 6}+\mu_{2,k_c}^{(d)} \right) \otimes \phi^3 + 
\nonumber \\[2mm]
&&\mu_{3,k_c}^{(e)} \otimes \phi^5 +\sum_{i=c,d,e} \eta^{(i)}_{k_c}
\, \dot{\phi } \;
\approx \; \sum_{N=1}^3 \xi_n \otimes \phi^{N-1}\;, \label{EOM2}
\end{eqnarray}
where $\otimes$ denotes a convolution in momentum space.  The coefficients
$\mu_i$ generally depend on {\bf k} and $t$, as well as on the momentum 
cut-off $k_c$.  Explicit calculation shows that the leading cut-off 
dependence of $\mu_1^k$ derives from the thermal one-loop contribution 
\cite{17} (see also \cite{LS95}):
\begin{equation}
\mu_{1,k_c}^{(a)} \, \longrightarrow \,
{\tilde g^2T^2\over 24} - {\tilde g^2k_cT\over 4\pi} \qquad 
\hbox{for $\tilde m=0$}\; . \label{oneloop}
\end{equation}
The vacuum contribution vanishes quadratically for small $k_c$.
The same is true for the one-loop correction to the coupling constant
\cite{LS95}:
\begin{equation}
\mu_{2,k_c}^{(d)} \, \stackrel{{\bf k} \rightarrow 0}{\approx} \,
-{3\tilde g^4T\over 16\pi m^*} \left(
1-{2k_c\over \pi m^*}\right)\;\buildrel\tilde m\to 0 \over \longrightarrow 
\;
{3\sqrt{6} \over 8\pi} \tilde g^3 \left( 1+ {3k_c\over \pi T}\right) - 
{9\tilde g^2 k_c \over \pi^2 T}\;, \label{loopcor}
\end{equation}
where ${m^*}^2=\tilde{m}^2+\mu_1(0)$ is the effective thermal mass.  As 
discussed
in \cite{17}, the cut-off dependence of $\mu_1$ is exactly balanced by the 
dynamically generated mass due to the self-interaction among the soft modes, 
if these are also in thermal equilibrium.  The same holds for the other 
$k_c$-dependent constants in (\ref{EOM2}).  If $\Gamma_{2N}^{k_c=0}$ contains
an infrared divergence, as often happens in massless quantum field theories, 
it is necessary to ``resum'' the thermal self-energy by including it into 
the propagator of the hard modes \cite{Par92,AE93}.  (See also \cite{12} 
for an extended discussion of this procedure.)

\section{Conclusions and Outlook}

In summary, we have succeeded in deriving a consistent set of
temporally local transport equations for the long-distance modes of a
self-coupled scalar field.  Our approach is based on
the analysis of time scales showing that the soft field modes, at high
temperature and weak coupling, oscillate on a time scale shorter than
the characteristic damping time.  A temporally local (Markovian)
equation can therefore only be derived for the Fourier components of
the soft field modes, i.e. the occupation amplitudes, but not for the
field itself.  In the spirit of quantum field theory, this corresponds
to the use of the interaction picture rather than the Heisenberg
picture.

As a result of our modified approach, we obtain finite values for the
dissipation coefficients without the need to introduce an explicit
damping for the hard field modes.  The dominant damping mechanism for
the infrared field mode $(k_c=0)$ is generally given by the two-loop
``sunset'' diagram, corresponding to emission or absorption of the
soft mode on hard thermal quanta.  This differs from the analysis of
Gleiser and Ramos \cite{12} who found that soft-mode damping is
dominated by scattering on hard particles in the traditional approach
of Morikawa \cite{9}.

How do our results generalize to other field theories?  Most
interesting quantum field theories, especially gauge theories, contain
three-particle vertices rather than, or in addition to, the four-point
coupling investigated here.  In these cases, there exists the standard
one-loop diagram, shown in Figure 3a for the $\phi^3$-theory.  The
real part associated to this
diagram has an imaginary part (in the $(k,\omega )$-representation)
in the space-like domain, corresponding
to the fluctuations in the mean field at space-like distances induced
by hard thermal particles.  While these fluctuations do not directly
produce damping of on-shell soft field modes, they provide a mechanism
for such damping, when coupled back into the nonlinear equation for
these modes.  Graphically, to lowest order,
this mechanism will generate a higher order diagram as shown in Figure 3b.
This corresponds, similar to our arguments presented in Appendix C,
to the imaginary part of the two loop diagram (also shown in Figure 3b)
when cut through the hard thermal loop and
the soft field line.  This analysis of mean field damping in the case
of three-point vertices is completely analogous to that performed by
Braaten and Pisarski \cite{BP90} for the damping of soft modes in
non-Abelian gauge theories.

Non-Abelian gauge fields at finite temperature would present an
interesting application of the methods discussed so far, because their
infrared sector remains non-perturbative even at high temperature due
to the absence of perturbative screening of magnetostatic fields. This 
phenomenon has been widely studied in the Euclidean formalism, and its 
resolution by a sequence of effective actions has been proposed 
\cite{B94,BN95}.  As mentioned in the Introduction, the situation is
more delicate in the real-time formalism because of the lack of a gap
in the excitation spectrum.

The emergence of a finite correlation length in the magnetostatic
sector is essentially a classical phenomenon which occurs at the momentum 
scale $g^2T$ \cite{Linde,Gross,Biro93,AY95}.  It should be fully contained 
in the classical effective action obtained by integrating out hard modes 
with momenta $\vert{\bf p}\vert > k_c > g^2T$.  In fact, our 
discussion in Section II.E suggests that we should choose $k_c \sim gT$ in 
non-Abelian gauge theories because the thermal gauge boson mass $m^*$ is 
of order $gT$.  This mass itself is of semiclassical origin 
\cite{Heinz,BI94,Kelly}, but requires the use of the Bose distribution 
for the occupation number of hard modes and hence is sensitive to quantum 
physics.

The problem in gauge theories is that one cannot employ a low-momentum 
cut-off $k_c$ because it is not gauge invariant.  A lattice formulation 
provides a gauge invariant cut-off through the lattice spacing $a$, 
roughly corresponding to $k_c =\pi/a$, but it violates rotational invariance.  
As pointed out in ref. \cite{17} this necessitates the introduction of 
rotationally non-invariant counterterms.  An infrared cut-off scheme that 
violates neither gauge nor Lorentz invariance is therefore preferable.  
Such a scheme, based on the proper time representation of the one-loop 
effective action, has recently been proposed \cite{Liao95}.

We found it convenient to formulate the effective equations of motion
(\ref{EOMb}) for the soft field modes in momentum space.  The question
arises whether these equations take a local form when written in
coordinate space.  This issue becomes important for gauge theories,
where the lattice regulated coordinate space representation appears
most natural.  In order to answer this question, it is necessary to
consider the momentum dependence of the kernels ${\cal M}^{(i)}(k,\omega)$
in (51).  In the extreme infrared limit $k_c\to 0$ only the term
${\cal M}^{(c)}$ survives.  The momentum dependence of this damping
coefficient $\eta^{(c)}$ has been analyzed by Jeon \cite{Je95} and by
Wang and Heinz \cite{EH95}, who found that it varies only weakly over the
range $0 \le \vert {\bf k}\vert < m^*$.  Outside that range, however, a strong
momentum dependence $\sim 1/|{\bf k}|$ is seen.  A complete analysis of the two 
other
damping coefficients $\eta^{(d)}$ and $\eta^{(e)}$ has not yet been
performed, but a strong momentum dependence appears likely.  On
the other hand, these terms are subdominant in the infrared domain.
Given that $E_k$ is also approaching a constant for $\vert {\bf
k}\vert \ll m^*$, it may be possible to obtain a quasi-local form of 
the equation for the soft modes at distances larger than $1/m^*$.
Clearly, a more complete analysis is required, also with regards
to the important issue whether the noise terms can be approximated as 
local white noise, before a detailed understanding of thermalization of 
soft modes in quantum field theory is achieved.  We hope to address 
these questions, as well as the issue of the numerical implementation of 
the equations derived here, in our future work.
\bigskip

\begin{center}
{\bf NOTE ADDED}
\end{center}
When finishing the manuscript the authors became aware of a recent 
manuscript by Boyanovsky, Lawrie and Lee on ``Relaxation and Kinetics in 
Scalar Field Theories'' \cite{BLL96}, where a linearized equation of motion
for the {\em ensemble} averaged field $\langle \Phi ({\bf x},t) \rangle$
is derived containing the contribution (a) and (c) of the real part of
the effective action (cf. Fig. 1). Their stated damping rate coincides
with our result (\ref{dampplasmon}).

\acknowledgments

We acknowledge stimulating discussions with M. Thoma.  C.G. thanks the 
Alexander von Humboldt Stiftung for its partial support with a Feodor 
Lynen scholarship and also acknowledges support by the BMBF and GSI 
Darmstadt.  This work was supported in part by the U.S. Department of
Energy under grants DE-FG05-90ER40592 and DE-FG02-96ER40945.

\newpage
\appendix

\section{Short review of the influence functional technique (Feynman)}

A quantum mechanical system X (described by the variable $x$)
interacts with a bath $Q$ (described by the variable $q$).  At some
specified initial time $t=t_i$ the combined system (X $\cup$ Q) is
described by the full density matrix
\begin{equation}
\rho_{{\rm X}\cup{\rm Q}}(t_i) = \rho_{{\rm X}\cup{\rm Q}}
(x_i,q_i,x_i',q_i';t_i). \label{A1}
\end{equation}
In the Schr\"odinger representation the (full) density matrix evolves
in time according to
\begin{eqnarray}
&&\rho_{{\rm X}\cup{\rm Q}}(x,q; x',q';t) \nonumber \\
&&\quad = \int dx_idq_idx'_idq'_i \left\{
U(x,q;x_i,q_i;t,t_i) \rho_{{\rm X}\cup{\rm Q}} (x_i,q_i;x'_i,q'_i;t_i)
U^{\dagger}(x'_i,q'_i;x',q';t,t_i)\right\}, \label{A2}
\end{eqnarray}
where the evolution operator reads in the path integral representation
\begin{equation}
U(x,q;x_i,q_i;t,t_i) = \int_{x_i}^x {\cal D}x \int_{q_i}^q {\cal D}q\; 
e^{{i\over\hbar}S[x,q]} \label{A3}
\end{equation}
with
\begin{equation}
S[x,q] = \int_{t_i}^t ds\; {\cal L}(x(s), q(s)). \label{A4}
\end{equation}
By formally integrating out the bath degrees of freedom $q$ one obtains
an effective interaction $S_{\rm eff} [x(s),x'(s)]$ describing the 
evolution of the system degrees of freedom $x(s)$.  This leads naturally 
to the ``doubling'' of the degrees of freedom in a real time description.

Typically one assumes that the initial density matrix $\rho_{{\rm X}\cup
{\rm Q}}(t_i)$ is uncorrelated in its variables $x$ and $q$, i.e.
\begin{equation}
\rho_{{\rm X}\cup{\rm Q}}(t_i) =
\rho_{\rm X}(x_i,x'_i,t_i) \otimes \rho_{\rm
Q}(q_i,q'_i;t_i), \label{A5}
\end{equation}
which can be motivated by assuming that the interaction 
${\cal L}_{\rm int}(x,q)$ is adiabatically switched on at the time $t_i$.

Introducing the reduced density matrix $\rho_r$ as
\begin{equation}
\rho_r(x,x';t) = {\rm Tr}_{(q)} \left\{ \rho(x,q;x',q';t)
\right\} \equiv \rho_{\rm X}(x,x';t) , \label{A6}
\end{equation}
one finds that the expectation value of any operator $\hat A$
depending solely on the system degrees of freedom can be readily
expressed in terms of $\rho_r$:
\begin{equation}
\langle A\rangle = {\rm Tr}_{(x,q)} \left\{ A(x,x';t)
\rho_{{\rm X}\cup{\rm Q}}(x,q;x',q';t)\right\} = {\rm Tr}_{(x)}
\left\{A(x,x';t) \rho_r(x,x';t)\right\}. \label{A7}
\end{equation}
Moreover one finds that its evolution in time can be put in the
general form
\begin{equation}
\rho_r(x,x';t) = \int dx_i dx'_i \int_{x_i}^x {\cal D}x \int_{x'_i}^{x'}
{\cal D}x'\; e^{{i\over\hbar}(S_x[x]-S_x[x'])}\;
e^{{i\over\hbar}S_{\rm IF}[x,x']} \rho_r(x_i,x'_i;t_i) \label{A8}
\end{equation}
by introducing the influence functional $S_{\rm IF}[x,x']$ \cite{21}
\begin{eqnarray}
e^{{i\over\hbar}S_{\rm IF}[x.x']} &= &{\rm Tr}_{(q)} \left\{
U_Q^{(x)}(q,q_i;t,t_i) \rho_Q(q_i,q'_i;t_i) U_Q^{(x') \,
\dagger }(q'_i,q_i;t,t_i)
\right\} \nonumber \\
&= &\int dq dq_i dq'_i \int_{q_i}^q {\cal D}q \int_{q'_i}^q 
{\cal D}q'\; e^{{i\over\hbar}(S_Q[q] - S_Q[q'] + S_{\rm int}[x,q] - 
S_{\rm int}[x',q'])} \rho_Q(q_i,q'_i;t_i). \label{A9}
\end{eqnarray}
Here $x(s), x'(s)$ are treated as given, classical background fields.  
Thus one is led to say that $\rho_r$ evolves in time according to the 
effective interaction
\begin{equation}
S_{\rm eff}[x,x'] = S_x[x] - S_x[x'] + S_{\rm IF}[x,x']. \label{A10}
\end{equation}
It is clear that $\rho_r(t)$ evolves causally from the history of the 
system and the bath.

A notational simplification is achieved by introducing combined
variables $x_c(s),q_c(s)$ defined on the real time contour ${\cal C}$
in the real time Green's function approach to finite temperature
quantum field theory or nonequilibrium quantum field theory, i.e.
\begin{equation}
x_c(s) = \cases{ x(s) &$s \in$ upper branch \cr
x'(s) &$s \in$ lower branch, \cr} \label{A11}
\end{equation}
and similarly for $q(s),q(s')$, as illustrated in Figure A1.  The
influence then takes the more compact form
\begin{equation}
e^{{i\over\hbar}S_{\rm IF}[x_c]} = \int dq dq_i dq'_i \int_{q_i}^q {\cal D}q
\int_{q'_i}^q {\cal D}q'\; e^{{i\over\hbar} \int^{\cal C} ds_c 
\left\{ {\cal L}_Q(q_c(s)) + {\cal L}_{\rm int}(x_c(s),q_c(s)) \right\}}
\rho_Q(q_i,q'_i;t_i), \label{A12}
\end{equation}
where the integration is defined on the time contour from $t_i$
forward to $t$ and back to $t_i$.

For the influence action $S_{\rm IF}[x,x']$ one finds the following
general properties to hold:
\begin{eqnarray}
S_{\rm IF}[x,x'] &= &-\left(S_{\rm IF}[x',x]\right)^*,  \label{A13} \\[2mm]
S_{\rm IF}[x,x] &= &0. \label{A14}
\end{eqnarray}
Expanding $S_{\rm IF}$ up to second order in $x$ and $x'$, the general
structure of $S_{\rm IF}$ is given by \cite{21}
\begin{eqnarray}
S_{\rm IF}[x,x'] &\approx &\int_{t_i}^t ds\; F(s) (x(s)-x'(s)) \nonumber \\
&+ &\frac{1}{2}\int_{t_i}^t ds_1ds_2 \left( x(s_1)-x'(s_1)\right) R(s_1,s_2)
\left( x(s_2)+x'(s_2)\right) \nonumber \\
&+ & \frac{i}{2} \int_{t_i}^t ds_1ds_2 \left( x(s_1)-x'(s_1)\right) I(s_1,s_2)
\left( x(s_2)-x'(s_2)\right) .\label{A15}
\end{eqnarray}
Suppose that ${\cal L}_{\rm int}$ takes the quite general form
\begin{equation}
{\cal L}_{\rm int}[x_c,q_c] = f \left( x_c(s)\right) \Xi
\left( q_c(s)\right). \label{A16}
\end{equation}
Writing for $S_{\rm IF}[x]$ up to second order in $f$
\begin{equation}
\exp{{i\over\hbar}S_{\rm IF}[x_c]} \approx e^{{i\over\hbar} \left[
\int^{\cal C} ds_1 F(s_1) f(x_c(s_1)) + {1\over 2} \int^{\cal C}
ds_1ds_2 f(x_c(s_1)) D(s_1,s_2) f(x_c(s_2))\right]} \label{17}
\end{equation}
it follows by comparing with (\ref{A12})
and performing the substitution $x(s) \rightarrow f(x(s))$
\begin{eqnarray}
F(s_1) &= &\left( {\hbar\over i}\right) {\delta e^{{i\over\hbar}S_{\rm IF}}
\over \delta f(x_c(s_1))} = \langle \Xi (q_c(s_1))
\rangle _{\rho _Q}  \nonumber \\
D(s_1,s_2) &= &\left( {\hbar\over i}\right) {\delta^2
e^{{i\over\hbar}S_{\rm IF}} \over \delta f(x_c(s_1))\delta f(x_1(s_2))} 
\nonumber \\
&= &{i\over\hbar} \left[ \langle P \left( \Xi (q_c(s_1)) \Xi
(q_c(s_2))\right) \rangle - \langle \Xi (q_c(s_1))\rangle \langle \Xi
(q_c(s_2))\rangle\right] \label{18}
\end{eqnarray}
where $P$ means path-ordering along the contour ${\cal C}$.  $F(s)$ can
be interpreted as an external force term due to the mean field generated
by the
average interaction of the bath variables $Q$ with the system  $X$.

The path-ordering definition leads to the typical four Green's functions
defined in real time:
\begin{eqnarray}
D^{++}(t_1,t_2) &= &{i\over\hbar} \left( \langle T (\Xi (q(t_1)) \Xi
(q(t_2))\rangle
\, - \,
\langle \Xi (q(t_1)) \rangle \langle \Xi (q(t_2)) \rangle \right)
\nonumber \\
&=& \theta(t_1-t_2) D^{-+}(t_1,t_2) + \theta(t_2-t_1)
D^{+-}(t_1,t_2) \nonumber \\
D^{+-} (t_1,t_2) &= &{i\over\hbar} \left(
\langle \Xi (q(t_2)) \Xi (q(t_1))\rangle
\, - \,
\langle \Xi (q(t_1)) \rangle \langle \Xi (q(t_2)) \rangle \right) \,
\equiv \, D^< (t_1,t_2) \nonumber \\
D^{-+}(t_1,t_2) &= &{i\over\hbar} \left(
\langle \Xi (q(t_1)) \Xi (q(t_2))\rangle
\, - \,
\langle \Xi (q(t_1)) \rangle \langle \Xi (q(t_2)) \rangle \right) \,
\equiv \, D^> (t_1,t_2) \nonumber \\
D^{--}(t_1,t_2) &= &{i\over\hbar} \left( \langle \tilde T (\Xi (q(t_1))
\Xi (q(t_2))\rangle
\, - \,
\langle \Xi (q(t_1)) \rangle \langle \Xi (q(t_2)) \rangle \right)
\nonumber \\
&=& \theta(t_1-t_2)D^{+-}(t_1,t_2) +
\theta(t_2-t_1) D^{-+}(t_2,t_2) \label{A19}
\end{eqnarray}
Here $T\;(\tilde T)$ stands for (anti)-time ordering.

Defining the {\em real} Green's functions
\begin{eqnarray}
D_R(t_1,t_2) &= &D^>(t_1,t_2) - D^<(t_1,t_2) = {i\over\hbar} \left\langle
\left[ \Xi (q(t_1)), \Xi (q(t_2))\right] \right\rangle \nonumber \\
D_I(t_1,t_2) &= &{1\over i} \left( D^>(t_1,t_2) + D^<(t_1,t_2)\right)
\nonumber \\
&=&{1\over\hbar} \Big( \left\langle \left\{ \Xi (q(t)1)), \Xi (q(t_2))
\right\}\right\rangle - 2 \left\langle\Xi (q(t_1))\right\rangle 
\left\langle \Xi (q(t_2))\right\rangle \Big) \, \, \, , \label{A20}
\end{eqnarray}
where
$D_R(t_1,t_2)=-D_R(t_2,t_1)$ and
$D_I(t_1,t_2)=D_I(t_2,t_1)$,
one finds after some algebra that in (\ref{A15}) $R$ and $I$ are given by
\begin{eqnarray}
R(t_1,t_2) &= & D_R(t_1,t_2) \theta(t_1-t_2) = 2{\rm Re}
(D^{++}(t_1,t_2)) \theta(t_1-t_2) \nonumber \\
I(t_1,t_2) &= &{1\over 2} D_I(t_1,t_2) =
{\rm Im}(D^{++}(t_1,t_2)).  \label{A21}
\end{eqnarray}
If the system behaves quasi-classically, the density matrix $\rho_r$
becomes nearly diagonal.  Then from (\ref{A8}), the major paths
contributing to the evolution of the reduced density matrix are obtained
by extremizing the effective action $S_{\rm eff}[x,x']$:
\begin{equation}
x(s) = x'(s)\quad \hbox{for all $t_i \le s < t$} \label{A22}
\end{equation}
and the extremity conditions
\begin{equation}
\left. {\delta S_{\rm eff}[x,x'] \over \delta x(s)} \right\vert_{x=x'}
= 0. \label{A23}
\end{equation}
The equations of motion are clearly causal if one constructs the effective 
action to the argument $s=t$.

Typically one introduces the variables
\begin{equation}
\bar x = {x+x'\over 2}, \quad \Delta = x-x' \label{A24}
\end{equation}
so that the semiclassical equations of motion (\ref{A23}) for $\bar x$
are stated as
\begin{equation}
\left. {\delta S_{\rm eff}(\bar x,\Delta) \over \delta\Delta(s)}
\right\vert_{\Delta=0} = 0. \label{A25}
\end{equation}
 From (\ref{A15}) we have by expansion up to second order
\begin{equation}
\frac{i}{\hbar } \, S_{\rm eff}(\bar x,\Delta) \, = \,
\frac{i}{\hbar} \, \Delta \otimes  \left( {\delta S_X[\bar x]
\over \delta \bar x(s)} + F(s) + R \otimes \bar x \right) \,
- \, \frac{1}{2\hbar }\Delta \otimes I \otimes\Delta, \label{A26}
\end{equation}
thus
\begin{equation}
-{\delta S_X[\bar x]\over \delta \bar x(s)} - F(s) - \int_{t_i}^s
ds' R(s,s') \bar x(s') = 0 \label{A27}
\end{equation}
as the {\em average} quasi-classical equation of motion.  This equation, in
return, has to be interpreted as an average over random, fluctuating
forces.  The last contribution in (\ref{A26}) leads to decoherence
because any path contributing with sizeable $\vert\Delta\vert >0$ over
past time becomes {\em exponentially} suppressed---hence, the imaginary part
$I$ drives the system to quasi-classical behavior.  For {\it short}
periods in time, however, fluctuations in $\Delta$ can appear on the
order of $\vert\int dt_1dt_2I(t_1,t_2)\vert^{-1/2}$ stochastically.  
These act as random ``kicks'' on the actual trajectory and can be 
interpreted as a stochastic force \cite{S82}.  To see this in
more detail, one defines the real stochastic influence action
\begin{equation}
\tilde{S}_{\rm IF}[x,x',\xi ] =
{\rm Re} (S_{\rm IF}[x,x']) + \int_{t_i}^t ds\;
\xi (s) \left(x(s)- x'(s)\right), \label{A28}
\end{equation}
where $\xi (s)$ is interpreted as an external force, randomly distributed
by a Gaussian distribution with zero mean:
\begin{equation}
P[\xi(s)] = {1\over N} \exp \left( -{1\over 2\hbar} \int_{t_i}^t ds_1ds_2
\xi(s_1) I^{-1}(s_1,s_2) \xi(s_2)\right). \label{A29}
\end{equation}
The influence functional $S_{\rm IF}[x,x']$  is regained as the
characteristic functional over the average of the random forces.
\begin{equation}
e^{iS_{\rm IF}[x,x']} \equiv \phi_{\xi} [x(s)-x'(s)] = \int D\xi \; 
P[\xi(s)]\; e^{i\tilde{S}_{\rm IF}[x,x',\xi]}. \label{A30}
\end{equation}
The imaginary part $I$ also defines the correlation of the random forces
\begin{equation}
\langle \xi(s_1) \xi(s_2)\rangle_{\xi} = \hbar I(s_1,s_2). \label{A31}
\end{equation}
 From (\ref{A28}) the corresponding equation of motion reads
\begin{equation}
-{\delta S_X[\bar x] \over \delta \bar x(s)} - F(s) -
\int_{t_i}^s ds' R(s,s') \bar x(s') = \xi(s) \label{A32}
\end{equation}
as the effective Langevin like equation dictating the dynamics of
the system variable in the quasi-classical regime.
(Classical) Brownian Motion \cite{S82}
is recovered if the spectral function of the system allows for approximating
the kernels $R$ and $I$ as
\begin{eqnarray}
R(s,s') & = & - m \gamma \dot{\delta } (s,s') \, \, \, ,
\nonumber \\
\hbar I(s,s') & = & 2m \gamma kT \delta (s,s') \, \, \ .
\label{A33}
\end{eqnarray}

In summary, by coarse-graining over the bath degrees of freedom,
there appear two major contributions to the evolution of the
reduced density matrix (\ref{A8},\ref{A26}), namely the oscillatory or
quantum mechanical phase and the decoherence factor when summing over
paths in the ($\bar x-\Delta$)-plane.  We do expect an essentially
semiclassical evolution when the excitation energy of the system 
(divided by $\hbar$) is large so that the oscillatory phases of paths 
neighboring the classical path interferes destructively.  Moreover, all
paths where the time averaged $\Delta$ is sufficiently large will be 
exponentially suppressed by the decoherence factor.  Hence, the system 
will be driven by decoherence into the quasi-classical regime, depending 
crucially on the magnitude $\sqrt{\langle \xi^2\rangle}\sim \sqrt{I}$ of the
random forces, i.e. the size of the noise kernel (\ref{A31}).  If it is
large, decoherence should happen very rapidly, and the quasi-classical 
description is inherently stochastic.

\section{The Influence Action for Scalar $\phi^4$ Theory}

Dividing the scalar field $\Phi$ into soft modes $\phi$ and hard modes
$\varphi$, the completely analogous expression for the influence 
functional can be read off (\ref{A12}):
\begin{equation}
e^{{i\over \hbar}S_{\rm IF}[\phi,\phi']} = \int d\varphi
\int_{\varphi_i}^\varphi {\cal D}\varphi \int_{\varphi'_i}^\varphi 
{\cal D}\varphi' e^{{i\over\hbar} \int^{\cal C} d^4s \left[ {\cal L}_h^0 
[\varphi] + {\cal L}'_{\rm int} [\phi,\varphi]\right]}
\rho_h (\varphi_i,\varphi'_i;t_i), \label{B1}
\end{equation}
where in the definition of $S'_{\rm int}[\phi,\varphi]$
\begin{equation}
S'_{\rm int}[\phi,\varphi] = g^2 \int_{t_i}^t d^4x  \left[
-{1\over 24}\varphi^4 - {1\over 6}\phi\varphi^3 - {1\over 4}
\phi^2\varphi^2 - {1\over 6}\phi^3\varphi\right] \label{B2}
\end{equation}
also the self interaction of the hard modes is included and
shall be treated perturbatively.

Expanding the influence functional up to order $g^4$,
\begin{equation}S_{\rm IF} = S_{\rm IF}^{(0)} + S_{\rm IF}^{(1)}
+S_{\rm IF}^{(2)}, \label{B3}
\end{equation}
one finds in an obvious notation:
\begin{eqnarray}
S_{\rm IF}^{(0)} &= &1, \nonumber \\
S_{\rm IF}^{(1)} &\equiv &\left\langle \int^{\cal C} d^4s 
{\cal L}'_{\rm int} [\phi,\varphi]\right\rangle_0 \equiv  \int d\varphi
\int_{\varphi_i}^\varphi {\cal D}\varphi \int_{\varphi_i}^{\varphi} 
{\cal D}\varphi' \left(\int^{\cal C} d^4s {\cal L}'_{\rm int}\right) 
e^{{i\over\hbar} \int^{\cal C} d^4s {\cal L}^0_h [\varphi ]}
\rho_h (\varphi_i,\varphi'_i;t_i)
\nonumber \\
S_{\rm IF}^{(2)} &= &{1\over 2}{i\over\hbar} 
\left( \left\langle \int^{\cal C} d^4s_1
d^4s_2 \; P\Big( {\cal L}'_{\rm int}(\phi(s_1),\varphi(s_1)) {\cal
L}'_{\rm int} (\phi(s_2),\varphi(s_2))\Big) \right\rangle_0 -
(S_{\rm IF}^{(1)})^2\right) \label{B4}
\end{eqnarray}
The expectation values can be evaluated perturbatively with standard
real time Green's function techniques \cite{Ch85,La87}.  To avoid
difficulties with initial higher order correlations of $\rho_h(t_i)$ one 
typically has to assume that $\rho_h(t_i)$ is of the form
\begin{equation}
\rho _{Q;0} \, \equiv \,
\rho_h(t_i) = \exp\left(\sum_p A_p\hat a^{\dagger}_p\hat a_p 
\right),\quad (A_p<0) \label{B5}
\end{equation}
to allow for a simple Wick decomposition.  A thermal distribution of
non-interacting hard particles is thus appropriate.  We define the real time
propagator of the hard particles on the contour as
\begin{equation}
G_c^{k_c}(x_1,x_2) = i \left\langle P \left( \varphi(x_1)\varphi(x_2)\right)
\right\rangle_0 \label{B6}
\end{equation}
with its components in real time
\begin{eqnarray}
G^{k_c}_{++}(x_1,x_2) &= &i\langle T(\varphi(x_1) \varphi(x_2))\rangle_0 =
\theta(t_1-t_2) G^{k_c}_{-+}(x_1,x_2) + \theta(t_2-t_1) G^{k_c}_{+-}(x_1,x_2)
\nonumber \\
G^{k_c}_{+-}(x_1,x_2) &= &i\langle\varphi(x_2) \varphi(x_1) \rangle_0 \equiv
G^{k_c}_<(x_1,x_2) \nonumber \\
G^{k_c}_{-+} (x_1,x_2) &= &i\langle\varphi(x_1)\varphi(x_2)\rangle_0 \equiv
G^{k_c}_>(x_1,x_2) \nonumber \\
G^{k_c}_{--}(x_1,x_2) &= &i
\langle \tilde T(\varphi(x_1)\varphi(x_2))\rangle_0,
\label{B7}
\end{eqnarray}
which have a simple Fourier representation
\begin{eqnarray}
G^{k_c}_>({\bf p},\tau) &= &G^{k_c}_<({\bf p},-\tau) = i
{[1+2n(\omega_p)]\cos\omega_p\tau - i\sin\omega_p\tau\over 2\omega_p}
\quad \hbox{for $\vert p\vert\ge k_c$} \nonumber \\
G^{k_c}_>({\bf p},\tau) &= &G^{k_c}_<({\bf p},-\tau) = 0 \quad \hbox{for
$\vert p\vert < k_c$} \label{B8}
\end{eqnarray}

The evaluation of the expectation values in (\ref{B4}) is lengthy but
straightforward. We state only the final results with their real and
imaginary parts:
\begin{eqnarray}
S_{\rm IF}^{(1a)} &= &i {g^2\over 4} \int^{\cal C} d^4x G_c^{k_c}(x,x)
\phi(x)^2  \nonumber \\
&= &i\; {g^2\over 4} \int_{t_i}^t d^4x
\left(\phi(x)-\phi'(x)\right) G^{k_c}_>(0)
\left(\phi(x)+\phi'(x)\right) \label{B9} \\
S_{\rm IF}^{(2b)} &= &-{g^4\over 8} \int^{\cal C}
d^4x_1 d^4x_2 \phi(x_1)^2 \left( G_c^{k_c}(x_1,x_2)\right)^2
G_c^{k_c}(x_2,x_2) \nonumber \\
&= &-{g^4\over 8} \int_{t_i}^t d^4x_1 d^4x_2
\left( \phi(x_1) -\phi'(x_1) \right) \nonumber \\
&& \quad \times \left\{ \theta(t_1-t_2)
\left( G^{k_c}_>(x_1,x_2)^2 - G^{k_c}_<(x_1,x_2)^2 \right)
G^{k_c}_>(0) \right\} \left( \phi(x_1) + \phi'(x_1) \right)
\label{B10} \\
S_{\rm IF}^{(2c)} & = &- {g^4\over 12} \int^{\cal C} d^4x_1d^4x_2 
\phi(x_1)
\left( G_c^{k_c}(x_1,x_2)\right)^3 \phi(x_2) \nonumber \\
&= &-{g^4\over 12} \int_{t_i}^t d^4x_1 d^4x_2 \left\{ (\phi(x_1) - 
\phi'(x_1))\theta(t_1-t_2) \right. \nonumber \\
&&\quad \times \left. \left( G^{k_c}_>(x_1,x_2)^3 - G^{k_c}_<(x_1,x_2)^3\right)
(\phi(x_2) + \phi'(x_2)) \right\} \nonumber \\
&&\quad  -i{g^4\over 12}  \int_{t_i}^t d^4x_1 d^4x_2 \left\{
(\phi(x_1)-\phi'(x_1)) \right. \nonumber \\
&&\quad \times \left. {1\over 2i} \left(G^{k_c}_>(x_1,x_2)^3 +
G^{k_c}_<(x_1,x_2)^3\right) (\phi(x_2)-\phi'(x_2)) \right\} \label{B11} \\
S_{\rm IF}^{(2d)} &= &-i{g^4\over 16} \int^{\cal C} d^4x_1 d^4x_2 
\phi (x_1)^2 \left( G_c^{k_c}(x_1,x_2)\right)^2 \phi (x_2)^2 \nonumber \\
&= &-i {g^4\over 16} \int_{t_i}^t d^4x_1 d^4x_2  \left\{ \left(
\phi(x_1)^2 - \phi'(x_1)^2\right) \theta(t_1-t_2) \right. \nonumber \\
&&\quad \times \left. \left( G^{k_c}_>(x_1,x_2)^2
- G^{k_c}_<(x_1,x_2)^2\right) \left(\phi(x_2)^2 + \phi'(x_2)^2\right)\right\}
\nonumber \\
&&\quad  -i{g^4\over 16}  \int_{t_i}^t d^4x_1 d^4x_2 \left\{ 
\left( \phi(x_1)^2 - \phi'(x_1)^2\right) \right. \nonumber \\
&&\quad \times \left. {1\over 2} \left( G^{k_c}_>(x_1,x_2)^2 +
G^{k_c}_<(x_1,x_2)^2\right) \left( \phi(x_2)^2 - \phi'(x_2)^2\right) \right\}
\label{B12} \\
S_{\rm IF}^{(2e)} &= &{g^4 \over 72} \int^{\cal C} d^4x_1d^4x_2 
\phi(x_1)^3 G_c^{k_c}(x_1,x_2) \phi(x_2)^3 \nonumber \\
&= &{g^4\over 72} \int_{t_i}^t d^4x_1 d^4x_2 \left\{ (\phi(x_1)^3 - 
\phi'(x_1)^3) \theta(t_1-t_2) \right. \nonumber \\
&&\quad \times \left. \left( G^{k_c}_>(x_1,x_2) - G^{k_c}_<(x_1,x_2)\right)
\left( \phi(x_2)^3 + \phi'(x_2)^3\right) \right\} \nonumber \\
&&\quad + i {g^4\over 72} \int_{t_i}^t d^4x_1 d^4x_2 \left\{ 
\left( \phi(x_1)^3 - \phi'(x_1)^3\right) \right. \nonumber \\
&&\quad \times \left. {1\over 2i} \left( G^{k_c}_>(x_1,x_2) +
G^{k_c}_<(x_1,x_2)\right) \left( \phi(x_2)^3 - \phi'(x_2)^3\right) \right\}
\label{B13}
\end{eqnarray}

Momentum space expressions can be found by inserting (\ref{B7}) into
these equations.  Introducing the new field variables
\begin{equation}
\bar\phi = {1\over 2} (\phi+\phi'), \qquad \phi_{\Delta} = \phi
-\phi' \label{B14}
\end{equation}
and using the relations
\begin{eqnarray}
{\rm Re}\left( G_{++}^{k_c}(x_1,x_2)^n\right) &= &\left(\theta(t_1-t_2)+(-1)^n
\theta(t_2-t_1)\right) {1\over 2} \left( G^{k_c}_>(x_1,x_2)^n + (-1)^n
G^{k_c}_<(x_1,x_2)^n \right) \nonumber \\
{\rm Im} \left( G_{++}^{k_c} (x_1,x_2)^n\right)
&= & \left( \theta(t_1-t_2)-(-1)^n
\theta(t_2-t_1)\right) {1\over 2i} \left( G^{k_c}_>(x_1,x_2)^n-(-1)^n
G^{k_c}_<(x_1,x_2)^n \right) \label{B15}
\end{eqnarray}
one finds after some calculation for the real and imaginary part,
respectively, of the influence action
\begin{eqnarray}
{\rm Re}\;&& S_{\rm IF}[\phi_{\Delta},\bar\phi] = -{g^2\over 2} 
\int_{t_i}^t d^4x_1 \phi_{\Delta}(x_1) \bar\phi(x_1)\;\; {\rm Im} \left[
G_{++}^{k_c}(0)\right] \nonumber \\
&&\quad +g^4 \int_{t_i}^t d^4x_1 d^4x_2 \theta(t_1-t_2) \left\{ 
{1\over 2} \phi_{\Delta}(x_1) \bar\phi(x_1) \; {\rm Im} G_{++}^{k_c}(0)\;
{\rm Im} \left[ G_{++}^{k_c}(x_1,x_2)^2\right] \right. \nonumber \\
&&\quad - {1\over 3} \phi_{\Delta}(x_1)\; {\rm Re}
\left[G_{++}^{k_c}(x_1,x_2)^3\right] \bar\phi(x_2)  \nonumber \\
&&\quad + {1\over 2} \bar\phi(x_1) \phi_{\Delta}(x_1)\; {\rm Im} \left[
G_{++}^{k_c}(x_1,x_2)^2\right] \left( \bar\phi(x_2)^2 + {1\over 4}
\phi_{\Delta}(x_2)^2\right) \nonumber \\
&&\quad \left. + {1\over 6} \phi_{\Delta}(x_1) \left( \bar\phi^2(x_1)+
{1\over 12}\phi_{\Delta}^2(x_1)\right)\; {\rm Re}
\left[G_{++}^{k_c}(x_1,x_2)\right] \bar\phi(x_2) \left( \bar\phi(x_2)^2 +
{3\over 4} \phi_{\Delta}(x_2)^2\right)
\right\}, \label{B16}
\end{eqnarray}
and
\begin{eqnarray}
{\rm Im}\; S_{\rm IF}\left[\phi_{\Delta},\bar\phi \right] & = &g^4
\int_{t_i}^t d^4x_1 d^4x_2 \left\{ -{1\over 12}\phi_{\Delta}(x_1)\; {\rm Im}
\left[G_{++}^{k_c}(x_1,x_2)^3\right] \phi_{\Delta}(x_2) \right. \nonumber \\
&&\quad -{1\over 4}\phi_{\Delta}(x_1) \bar\phi(x_1)\; {\rm Re }\left[
G_{++}^{k_c}(x_1,x_2)^2\right] \phi_{\Delta}(x_2) \bar\phi(x_2) \nonumber \\
&&\quad + {1\over 8} \phi_{\Delta}(x_1) \left( \bar\phi(x_1)^2 +
{1\over 12} \phi_{\Delta}(x_1)^2\right)\nonumber \\
&&\quad \times {\rm Im} \left[ 
G_{++}^{k_c}(x_1,x_2)\right] \phi_{\Delta}(x_2) \left( \bar\phi (x_2)^2
+ {1\over 12} \phi_{\Delta}(x_2)^2\right). \label{B17}
\end{eqnarray}
After introducing the stochastic auxiliary functions $\xi_i(x),\;
i=1,2,3$, the imaginary part of the influence action becomes
\begin{eqnarray}
&&{\rm Im}\; S_{\rm IF}[\phi_{\Delta},\bar\phi] \to \tilde S_{\rm IF} \left[
\bar\phi,\phi_{\Delta}; \xi_i\right] \nonumber \\
&&\qquad = \int_{t_i}^t d^4x 
\left[ \phi_{\Delta}(x) \xi_1(x) + \phi_{\Delta}(x) \bar\phi(x)
\xi_2(x) +\phi_{\Delta}(x) \left( \bar\phi(x)^2 + { 1\over 12}
\phi_{\Delta}(x)^2 \right) \xi_3(x) \right] \label{B18}
\end{eqnarray}
The corresponding stochastic weights are
\begin{eqnarray}
P[\xi_1] &= &N_1 \left\{ \int_{t_i}^t d^4x_1 d^4x_2 \xi_1(x_1) 
{3\over g^4} \left[ {\rm Im} \left( G_{++}^{k_c} \right)^3
\right]_{x_1,x_2}^{-1} \xi_1(x_2) \right\} \nonumber \\
P[\xi_2] &= &N_2 \left\{ \int_{t_i}^t d^4x_1d^4x_2 \xi_2(x_1) 
{1\over g^4} \left[ {\rm Re} \left( G_{++}^{k}\right)^2
\right]_{x_1,x_2}^{-1} \xi_2(x_2) \right\}  \nonumber \\
P[\xi_3] &= &N_3 \left\{ \int_{t_i}^t d^4x_1 d^4x_2 \xi_3(x_1) \left(
{-2\over g^4}\right) \left[ {\rm Im} \left( G_{++}^{k_c}\right)
\right]_{x_1,x_2}^{-1} \xi_3(x_2) \right\}. \label{B19}
\end{eqnarray}

\section {Derivation of two Damping Coefficients}

Following the interpretation developed in Section III.A, the dissipative 
contribution due to ${\rm Re}\left( S_{\rm IF}^{(d)} \right)$ to the 
equation of motion of the low momentum mode $\phi ({\bf k},t)$ is stated as:
\begin{eqnarray}
&&\int^{k_c} \frac{d^3k_2}{(2\pi )^3} \,
\Upsilon^{(d)}_{\rm diss.}({\bf k}-{\bf k}_2,t) \,
\phi ({\bf k}_2,t) = \int^{k_c} \frac{d^3k_2}{(2\pi)^3} \,
\Upsilon^{(d)}_{\rm diss.}({\bf k}+{\bf k}_2,t)
\phi^* ({\bf k}_2,t) \nonumber \\
&&\quad = \int^{k_c} \frac{d^3k_1d^3k_2}{(2\pi)^6} 
\, \dot{\phi }({\bf k}-{\bf k}_1+{\bf k}_2,t) \,
\phi ({\bf k}_1,t) \phi^* ({\bf k}_2,t)\;
\frac{1}{4 E_{{\bf k}-{\bf k}_1 + {\bf k}_2} }
\, \theta (k_c-|{\bf k}-{\bf k}_1+{\bf k}_2|) \nonumber \\
&& \qquad \times \left(
i{\cal M}^{(d)}({\bf k}+{\bf k}_2,E_{{\bf k}-{\bf k}_1+{\bf k}_2}+E_{k_1}) +
i{\cal M}^{(d)}({\bf k}+{\bf k}_2,E_{{\bf k}-{\bf k}_1+{\bf k}_2}-E_{k_1})
\right) \; .  \label{C1}
\end{eqnarray}
This is a highly nonlinear term where the integrand rapidly oscillates
over the whole integration region. However, its `major' contribution
is expected for momenta ${\bf k}_1 \approx {\bf k}_2$ because then
$\phi ({\bf k}_1,t) \phi^* ({\bf k}_2,t) \approx |\phi ({\bf k}_1,t) 
|^2 >0$.  It is not our objective to show that all other contributions 
cancel, but to extract the dominant contribution to the integrals in 
(\ref{C1}).  Indeed, setting ${\bf k}_1={\bf k}_2$ corresponds to the 
Wick contraction in standard thermal field theory.  Suppose the above 
expression would be taken as an ensemble average over all possible and 
similar configurations, i.e. initial conditions, thus averaging over all 
the fluctuating phases.  For a quasi-homogeneous system one will find for 
the correlator 
\begin{equation}
\langle \phi ({\bf k}_1,t) \phi^* ({\bf k}_2,t) \rangle
\approx (2\pi )^3 \delta ^3({\bf k}_2 -{\bf k}_1) \,
|\phi ({\bf k}_1,t) |^2 \; . \label{C2}
\end{equation}
Here
\begin{equation}
|\phi ({\bf k}_1,t) |^2 \approx \frac{1}{E_{k_1}} \, N({\bf k}_1,t)
\approx \frac{1}{E_{k_1}} \, (1+N({\bf k}_1,t))\; , \label{C3}
\end{equation}
which is identified with the average population number in the classical 
regime, when the mode population is large.  (\ref{C2}) is then completely 
analogous to the equal-time limit of the propagators (defined in Appendix B) 
in a homogeneous system:
\begin{equation}
-iG_>({\bf k},\tau =0) \approx  n(\omega_k)/ \omega_k \;. \label{C4}
\end{equation}
As a further motivation let us first calculate the classical
vertex contribution to the equation of motion by means of
the approximation (\ref{C2}):
\begin{eqnarray}
&&\frac{\tilde{g}^2}{3!} \int^{k_c}\frac{d^3k_1d^3k_2}{(2\pi )^6}
\, \phi ({\bf k}-{\bf k}_1-{\bf k}_2,t)
\phi ({\bf k}_1,t) \phi ({\bf k}_2,t)
\, \theta (k_c-|{\bf k}-{\bf k}_1-{\bf k}_2|) \nonumber \\
&&\qquad \approx 
\left( \frac{\tilde g^2}{2} \int^{k_c} \frac{d^3k_1}{(2\pi )^3}
\, |\phi ({\bf k}_1,t) |^2 \right) \, \phi ({\bf k},t)
\, \equiv \, \Delta m^2_{{\rm cl}} \, \phi ({\bf k},t) \; . \label{C5}
\end{eqnarray}
Neglecting the fluctuations let us write the average influence of the
other soft modes as a mass term induced by the mean field. This mass term
corresponds to the lowest order tadpole contribution in the quantum field 
theory.  Noticing the identification (\ref{C3}), we find that 
$\Delta m_{\rm cl}^2$ exactly equals the expression (\ref{mgap}) 
stated in the Introduction.  Now, writing
\begin{equation}
\dot{\phi}({\bf k}-{\bf k}_1+{\bf k}_2,t) \,
\phi ({\bf k}_1,t) \phi^* ({\bf k}_2,t) = 
\dot{\phi }({\bf k},t) \,\langle \phi ({\bf k}_1,t) 
\phi^* ({\bf k}_2,t) \rangle + {\rm fluctuations} \; , \label{C6}
\end{equation}
the first term contributes to (\ref{C1}) as:
\begin{eqnarray}
\dot{\phi }({\bf k},t) \frac{1}{4 E_{{\bf k}} }
\int^{k_c} \frac{d^3k_1}{(2\pi )^3} \, |\phi ({\bf k}_1,t)|^2 \,
&&\left( 
i{\cal M}^{(d)}({\bf k}+{\bf k}_1,E_{{\bf k}}+E_{k_1}) +
i{\cal M}^{(d)}({\bf k}+{\bf k}_1,E_{{\bf k}}-E_{k_1}) \right)
\nonumber \\
&&\equiv \eta ^{(d)}({\bf k},t) \, \dot{\phi }({\bf k},t) \;. \label{C7}
\end{eqnarray}
With the identification (\ref{C3}) this expression thus corresponds
naturally to the scattering diagram obtained by cutting the `sunset' 
diagram with two hard and one soft propagator as illustrated in Figure C1.

Similarly, for the dissipative contribution of 
${\rm Re}\left( S_{\rm IF}^{(e)} \right)$ we write for the first term:
\begin{eqnarray}
&&\int^{k_c} \frac{d^3k_1'd^3k'_2d^3k_1d^3k_2}{(2\pi )^{12}}
\dot{\phi }({\bf k}_3,t) \, \phi ({\bf k}_1,t) \phi ({\bf k}_2,t)
\phi^* ({\bf k}_1',t) \phi^* ({\bf k}_2',t) \,
\theta (k_c-|{\bf k}_3|) \nonumber \\
&& \quad \times \frac{3}{8E_3} 
\left( i{\cal M}^{(e)}({\bf k}+{\bf k}_1'+{\bf k}_2',E_1+E_2+E_3) +
i{\cal M}^{(e)}({\bf k}+{\bf k}_1'+{\bf k}_2',-E_1-E_2+E_3) \right.
\nonumber \\
&&\qquad \left.  + 2i{\cal M}^{(e)}({\bf k}+{\bf k}_1'+{\bf k}_2',
E_1-E_2+E_3) \right)  \label{C8} 
\end{eqnarray}
with ${\bf k}_3 = {\bf k} + {\bf k}_1' + {\bf k}_2' -{\bf k}_1 
-{\bf k}_2$.  The other term is of higher order in the time derivatives 
and can be neglected if the mode population changes weakly in time, i.e.
$$
\frac{d}{dt} |\phi ({\bf k},t)|^2 \, \ll \, E_k |\phi ({\bf k},t)|^2\; .
$$
Again, the dominant contribution to the integral arises when
${\bf k}_1\approx {\bf k}_1', \, {\bf k}_2\approx {\bf k}_2'$ or
${\bf k}_1\approx {\bf k}_2', \, {\bf k}_2\approx {\bf k}_1'$, so that
we approximate for a quasi-homogeneous system:
\begin{eqnarray}
&&
\langle \phi ({\bf k}_1,t) \phi ({\bf k}_2,t)
\phi^* ({\bf k}_1',t)
\phi^* ({\bf k}_2',t)
\rangle
\nonumber \\[2mm]
\approx &&
(2\pi )^6 \left( \delta ^3({\bf k}_1 -{\bf k}_1')
\delta ^3({\bf k}_2 -{\bf k}_2')
\, + \, \delta ^3({\bf k}_1 -{\bf k}_2')
\delta ^3({\bf k}_2 -{\bf k}_1') \right)
\,
|\phi ({\bf k}_1,t) |^2
|\phi ({\bf k}_2,t) |^2
\label{C9}
\end{eqnarray}
Inserting this into (\ref{C8}) one obtains
\begin{eqnarray}
&&\dot{\phi }({\bf k},t) \,
\frac{3}{4E_k} \int^{k_c} \frac{d^3k_1d^3k_2}{(2\pi)^6}\,
|\phi ({\bf k}_1,t) |^2 \,|\phi ({\bf k}_2,t) |^2 \nonumber \\
&&\quad \times \left(
i{\cal M}^{(e)}({\bf k}+{\bf k}_1+{\bf k}_2,E_1+E_2+E_k) +
i{\cal M}^{(e)}({\bf k}+{\bf k}_1+{\bf k}_2,-E_1-E_2+E_k) \right. \nonumber \\
&&\qquad + \left. 2i{\cal M}^{(e)}({\bf k}+{\bf k}_1+{\bf k}_2,E_1-E_2+E_k) 
\right) \nonumber \\
\equiv && \eta^{(e)}({\bf k},t) \,
\dot{\phi }({\bf k},t) \; . \label{C10}
\end{eqnarray}
This corresponds to cutting the `sunset' diagram with two internal soft
propagators and one hard propagator giving rise to the various on-shell 
scattering and production processes as illustrated in Figure C2.

\input epsf

\begin{figure}
\def\epsfsize#1#2{.90#1}
\centerline{\epsfbox{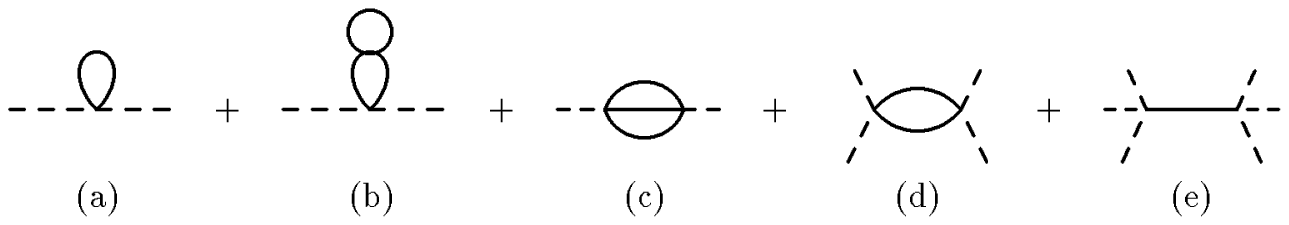}}
{\footnotesize 
FIG.1. Feynman diagrams contributing to the influence action up to
order $g^4$. Full lines denote hard modes and dashed lines correspond
to soft modes.}
\label{fig1.ps}
\end{figure}

\begin{figure}
\centerline{\epsfbox{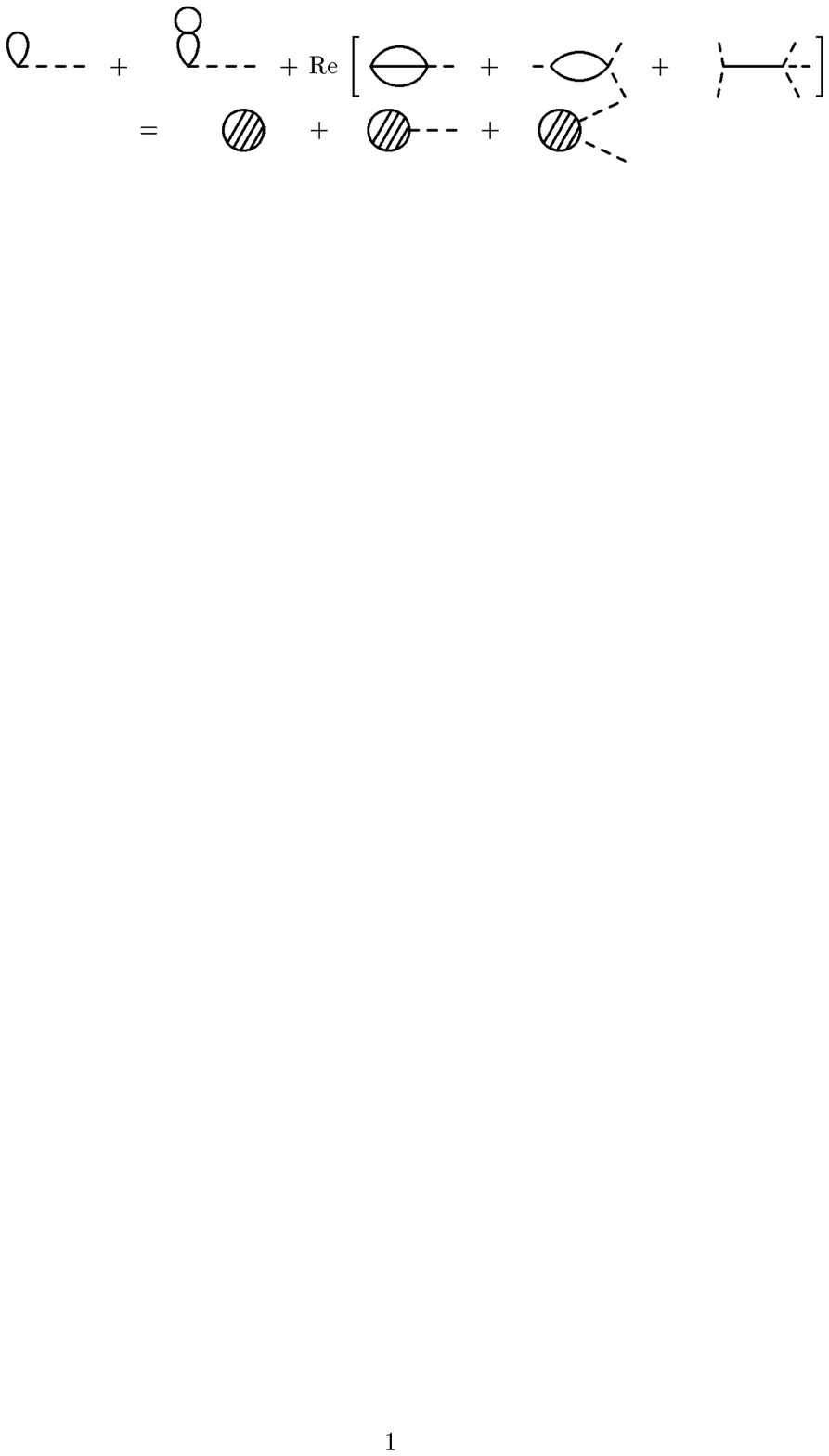}}
{\footnotesize 
FIG.2. Graphical representation of the classical equation
(\protect\ref{EOM}) for the soft field modes.  The noise terms 
$\phi^{N-1}\xi_N$ are shown as blobs with $(N-1)$ external legs.}
\label{fig2.ps}
\end{figure}

\begin{figure}
\centerline{\epsfbox{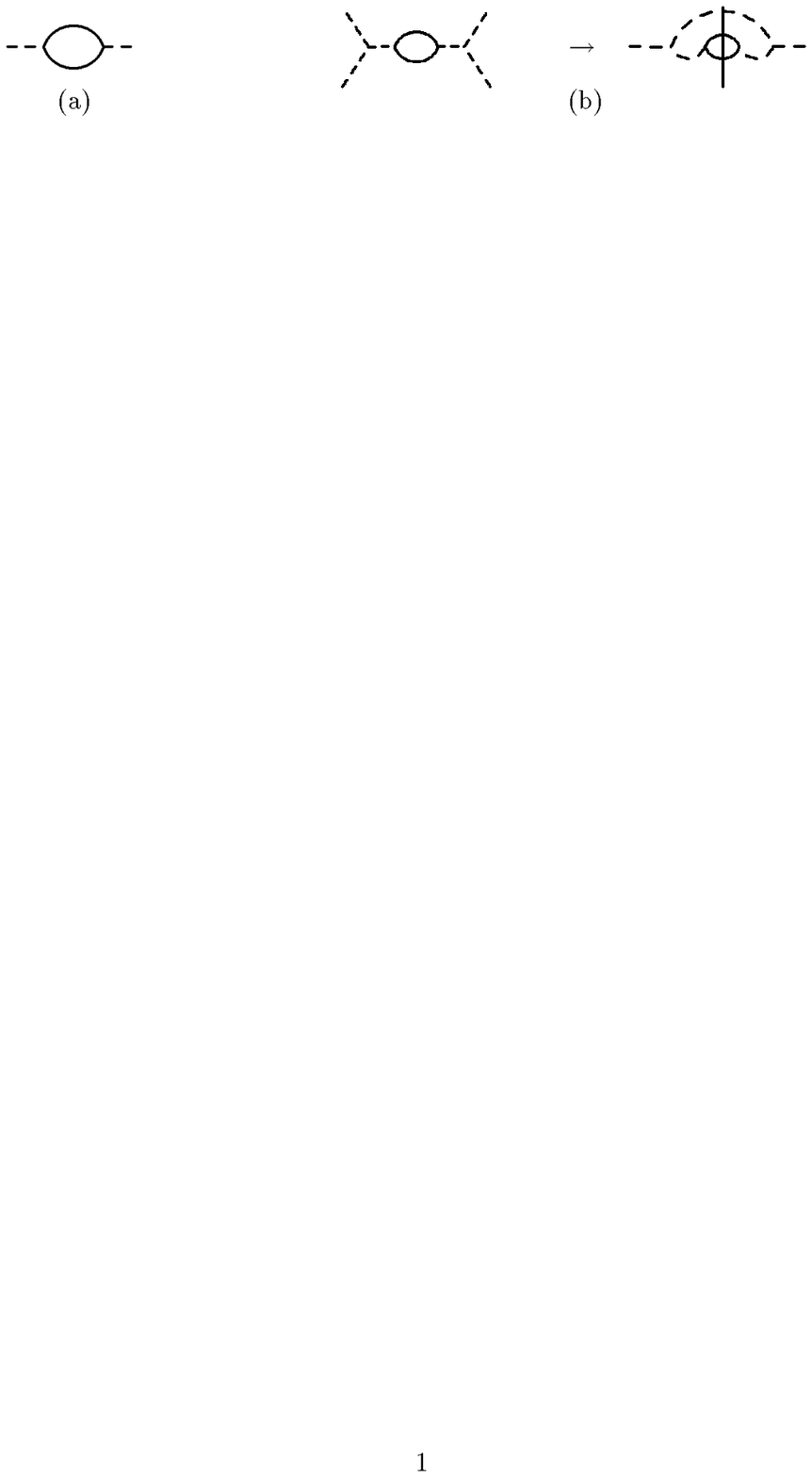}}
{\footnotesize 
FIG.3. (a) A particular contribution to the effective action at order
$g^2$ in $\Phi ^3$-theory;
(b) Lowest order diagram leading to on-shell dissipation
when iterating the real part ${\rm Re}(S_{IF}^{(a)})$ of (a) together
with the $\frac{g}{3!}\phi^3$ term in the full nonlinear and semiclassical
equation of motion. It corresponds to the usual two loop diagram
when cut through the hard thermal loop and the soft (thermal) field line.}
\label{fig3.ps}
\end{figure}

\vfill\eject

\begin{figure}
\centerline{\epsfbox{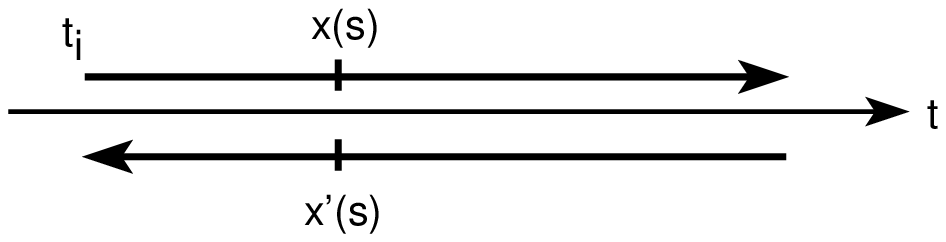}}
{\footnotesize  
FIG.A1. Schwinger-Keldysh time contour path ${\cal C}$
for the variables $x(s)$ and $x'(s)$ running from $s=t_i$ to $+\infty$
and back to $t_i$.}
\end{figure}

\begin{figure}
\centerline{\epsfbox{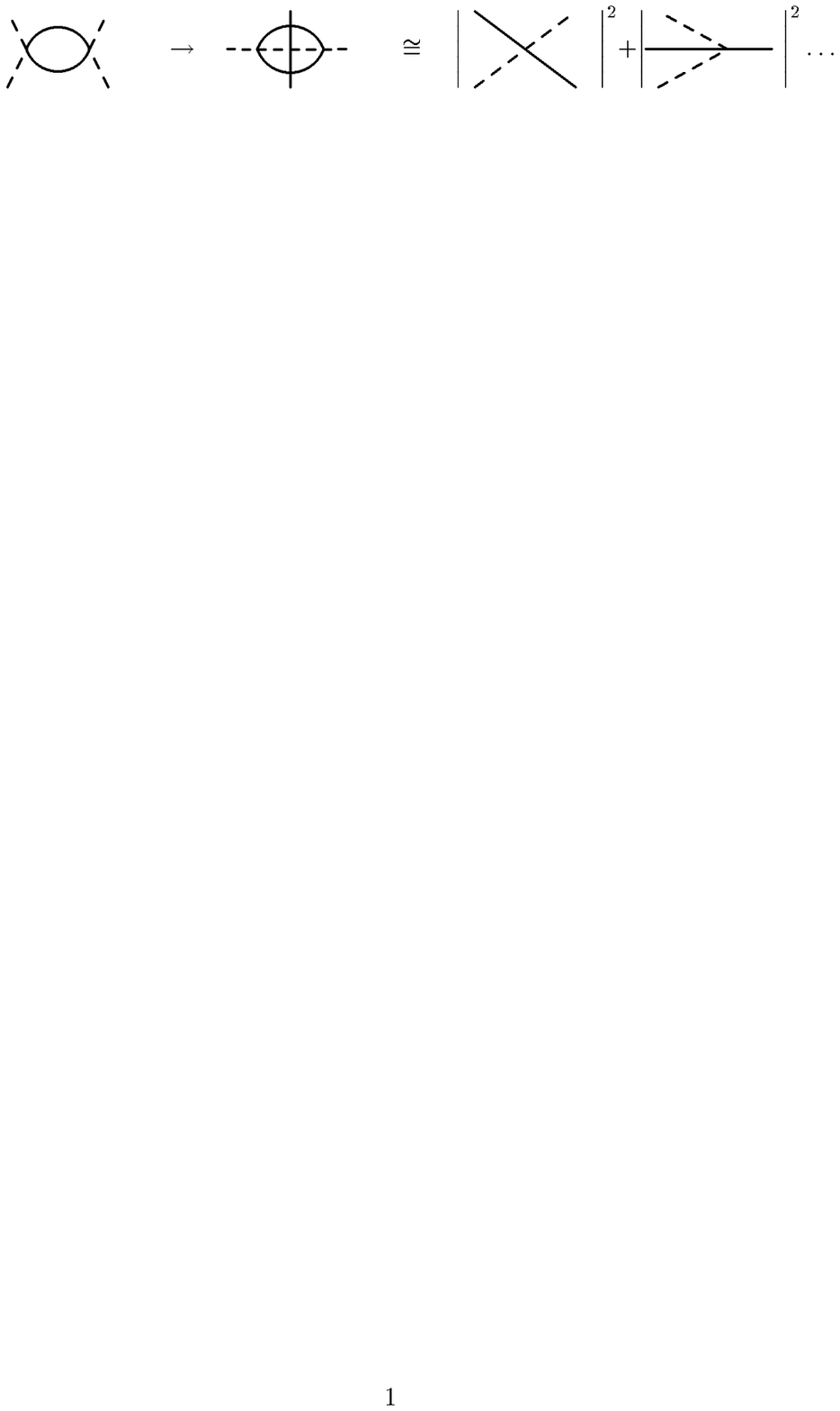}}
{\footnotesize  
FIG.C1. Various scattering contributions when cutting
the ``sunset'' diagramm with two hard particles and one soft mode.}
\end{figure}

\begin{figure}
\centerline{\epsfbox{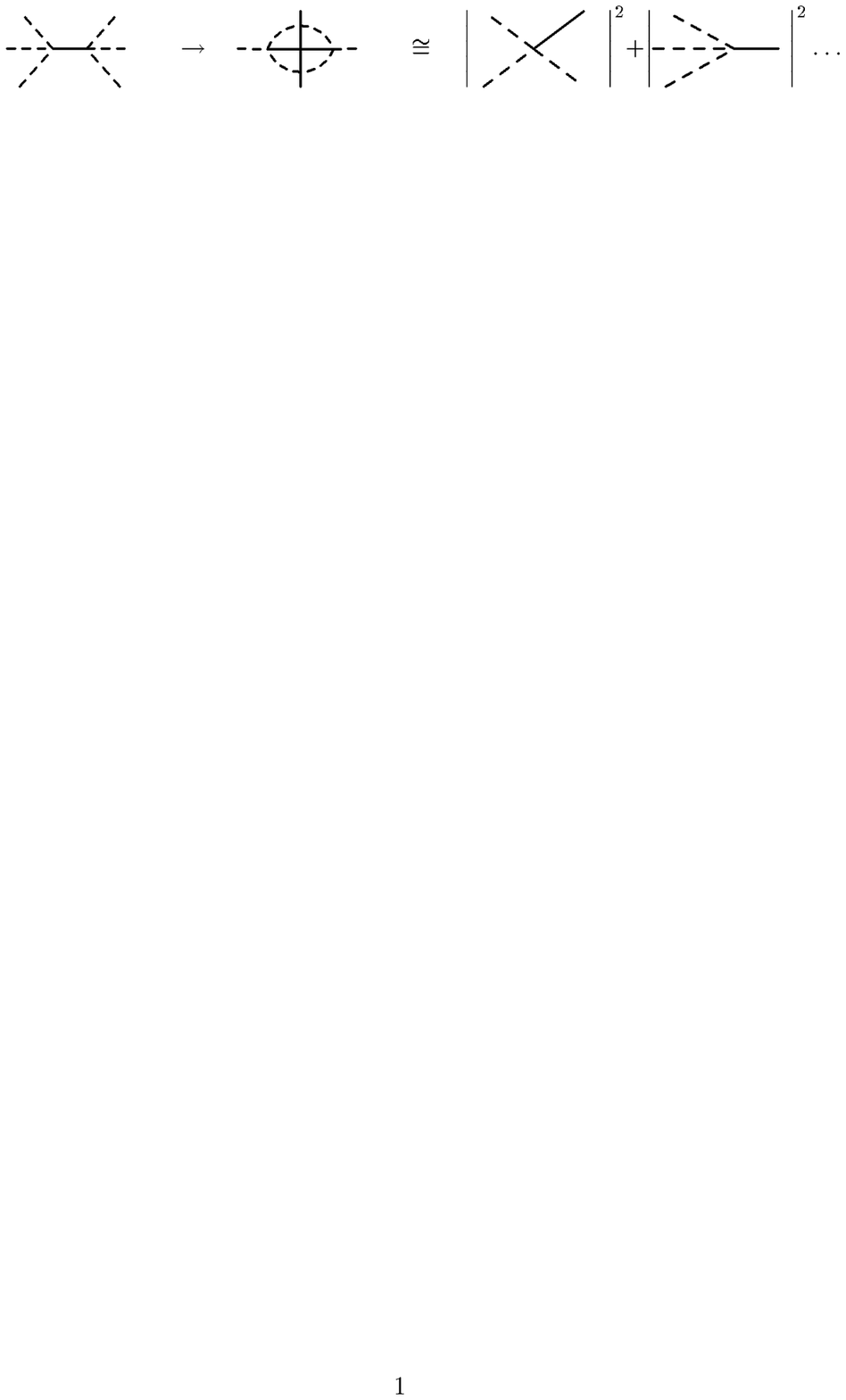}}
{\footnotesize  
FIG.C2. Various scattering contributions when cutting
the `sunset' diagram with one hard particle and two soft modes.}
\end{figure}

\end{document}